\documentclass[10pt,aps,prb,twocolumn,superscriptaddress,amsmath,amssymb,floatfix,longbibliography]{revtex4-2}
\usepackage{graphicx}
\usepackage{bm}
\usepackage{booktabs}
\usepackage{mathtools}
\usepackage{placeins}
\usepackage[hidelinks]{hyperref}
\newcommand{\Arg}{\operatorname{Arg}}
\newcommand{\sgn}{\operatorname{sgn}}

\renewcommand{\dbltopfraction}{0.95}

\newbox\appcolsbox
\newbox\appleftbox
\newdimen\appcolwidth
\newdimen\apphalfheight
\newcommand{\appcols}[1]{\begingroup
\appcolwidth=\dimexpr(\linewidth-\columnsep)/2\relax
\setbox\appcolsbox=\vbox{\hsize=\appcolwidth\linewidth=\hsize\columnwidth=\hsize\parskip=0pt #1\par}
\apphalfheight=\ht\appcolsbox\advance\apphalfheight by\dp\appcolsbox\divide\apphalfheight by2\advance\apphalfheight by6pt
\setbox\appleftbox=\vsplit\appcolsbox to\apphalfheight
\noindent\hbox to\linewidth{\vtop{\hsize=\appcolwidth\unvbox\appleftbox}\hfil\vtop{\hsize=\appcolwidth\unvbox\appcolsbox}}\par
\endgroup}
\begin{document}
\title{Local Josephson response near nonmagnetic impurities and vortices in a two-orbital \texorpdfstring{$s_{\pm}$}{s+-} superconductor}
\author{Yuanyuan Zhao}
\email{zhaoyuanyuan@gmail.com}
\affiliation{Key Laboratory of Intelligent Optoelectronic Devices and Chips of Jiangsu Higher Education Institutions, School of Physical Science and Technology, Suzhou University of Science and Technology, Suzhou, 215009, China}
\affiliation{Quantum Science Center of Guangdong-Hong Kong-Macao Greater Bay Area, Shenzhen 518045, China}
\begin{abstract}
We study the local Josephson response near nonmagnetic impurities and vortices in a self-consistent two-orbital $s_{\pm}$ superconductor. For a local $s$-wave tip that couples equally to the two orbitals, anomalous spectral contributions of opposite sign nearly cancel in the clean sample. Nonmagnetic impurities redistribute these contributions and can reverse the local coupling without a corresponding $\pi$ phase shift of the bond pair fields. Negative coupling appears at off-center sites at weaker impurity potentials than at the impurity site. The impurity-induced change is concentrated at lower quasiparticle energies, while states over a broader energy range contribute to the total clean and impurity couplings. In zero field, the coupling also changes sign as the tip gap is varied, and increasing the tunneling strength can reverse the $m=1$ sine component of the current without changing the sample state. At the chosen coherence-peak energy, increasing impurity strength mainly changes the relative intensities of existing quasiparticle-interference features. In a self-consistent two-vortex state, the Josephson response develops a spatially varying phase relative to the condensate. The response and condensate share the winding close to each core, while larger loops enclose additional winding of the Josephson response. The local Josephson sign and phase reflect the anomalous spectrum and the tip--sample coupling.
\end{abstract}
\maketitle
\setlength{\parskip}{0pt}
\flushbottom
\predisplaypenalty=0

\makeatletter
\def\section{\@startsection{section}{1}{\z@}{6pt}{4pt}{\normalfont\small\bfseries\centering}}
\def\subsection{\@startsection{subsection}{2}{\z@}{4pt}{3pt}{\normalfont\small\bfseries\centering}}
\makeatother
\setlength{\abovedisplayskip}{8pt}
\setlength{\belowdisplayskip}{8pt}
\setlength{\abovedisplayshortskip}{4pt}
\setlength{\belowdisplayshortskip}{4pt}

\section{Introduction}
\label{sec:intro}

Josephson scanning tunneling microscopy (JSTM) uses a superconducting tip to probe local pair tunneling \cite{Smakov2001} and has enabled measurements with atomic spatial resolution \cite{Randeria2016,Hamidian2016,Cho2019}. Ordinary tunneling spectroscopy probes the electron local density of states (LDOS), whereas the Josephson current depends on anomalous pair correlations in the tip and sample. The tunneling matrix elements determine which sample correlations contribute to the current. Interpreting a Josephson map therefore requires understanding how this local pair-tunneling response is related to the sample pair field.

Depending on the tunneling geometry, the Josephson current can follow the spatial variation of the sample pair field near a defect \cite{Graham2017}. In the disordered unconventional
superconductors studied in Ref.~\cite{Sulangi2021}, the pair field and
Josephson critical current remain closely related even when their correlation
with the spectroscopic gap is weak. Pairing symmetry and tunneling matrix
elements can also produce a different relation: an $s$-wave tip on a $d$-wave
surface can yield vanishing local Josephson coupling and spatial sign reversals
\cite{Choubey2024}. 

In multiband $s_{\pm}$ superconductors, pair contributions of opposite sign
can enter the same Josephson contact. Their competition can produce small net
currents, sign changes and enhanced higher harmonics near compensation
\cite{Agterberg2002Multigap,Ota2009Multigap,Ota2010Multigap,
Sperstad2009Transport,Koshelev2012}. Josephson scanning tunneling measurements on FeSe
have examined frustrated coupling and the tuning of relative tunneling-channel
transparencies \cite{Sharma2025FeSe}. For a spatially nonuniform sample, the
balance also depends on the quasiparticle wave functions at the contact. A local defect changes the quasiparticle states and the self-consistent pair field, thereby redistributing the anomalous spectral contributions to the local Josephson coupling.

We study the local Josephson response near nonmagnetic impurities and vortices in a self-consistent two-orbital $s_{\pm}$ superconductor, using a local $s$-wave tip with equal orbital weights. We examine how the sign and phase of this response are related to the sample bond pair field. The clean local coupling is already small because anomalous spectral contributions of opposite sign nearly cancel.
Resolving the Josephson coupling over the Bogoliubov--de Gennes (BdG) spectrum identifies which sample energies control the impurity-induced change and which contribute to the total Josephson coupling. Changing the tip gap changes the relative contributions of sample states at different energies to pair tunneling. We also calculate
current--phase relations (CPRs) beyond the weak-tunneling limit for fixed sample fields to
examine how the $m=1$ Josephson harmonic changes with tunneling strength.
Determining the sign of the $m=1$ sine coefficient experimentally requires a phase reference \cite{Karan2022}. We distinguish the signed sine coefficient of the $m=1$ harmonic from the nonnegative critical-current amplitude.

Quasiparticle interference (QPI) probes scattering in the same self-consistent impurity states \cite{Hoffman2002,WangLee2003,Zhang2009QPI,
Akbari2013FullT,Hong2021}. Its scattering features arise from impurity modulation of the electron LDOS in the superconducting state and depend on superconducting coherence factors, while the Josephson coupling is determined by anomalous propagation. For the same self-consistent impurity states, we compare the impurity-induced LDOS modulations with the changes in local pair tunneling. For the self-consistent two-vortex state, we compare the magnitude, phase and winding of the local Josephson response with those of the local bond pair field. Vortex-core spectra have been calculated in multiorbital models \cite{Hu2009Vortex,Gao2011Vortex}, and pair tunneling around a vortex has been measured with a superconducting tip \cite{Bergeal2008}.

The model and local observables are introduced in Sec.~\ref{sec:method}. Sections~\ref{sec:josephsonresults}--\ref{sec:contactresults} examine the nonmagnetic-impurity response, its spectral contributions, and its dependence on the tip gap and tunneling strength. QPI is considered in Sec.~\ref{sec:normal}, followed by the magnitude, phase and winding of the Josephson response in the two-vortex state in Sec.~\ref{sec:vortices}. Sections~\ref{sec:discussion} and \ref{sec:conclusion} discuss and summarize the results.

\begin{figure*}[!t]
\centering
\includegraphics[width=430bp]{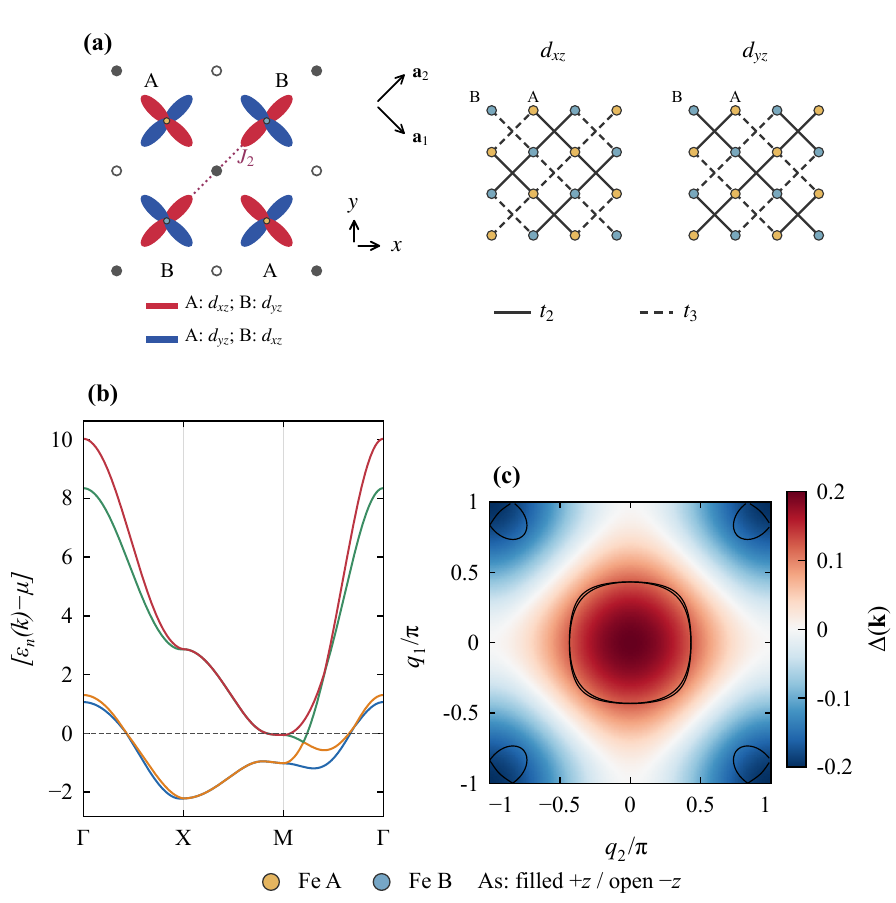}
\caption{\label{fig:model}Model and clean superconducting state.
(a) Four Fe sites, surrounding As positions and a representative next-nearest-neighbor (NNN) pair bond $J_2$. The alternating As heights depict the idealized Fe--As coordination environment with $D_{2d}$ symmetry. Only the two Fe orbitals are retained explicitly; the As environment is represented through effective Fe--Fe hoppings. For related symmetry-based two-orbital modeling, see Ref.~\cite{HuHao2012S4}. Red denotes $d_{xz}$ on A and $d_{yz}$ on B; blue denotes $d_{yz}$ on A and $d_{xz}$ on B. The two right-hand lattice patches retain the fixed $d_{xz}/d_{yz}$ labels used in the calculation. Black diagonal links connect NNN Fe sites within the same sublattice. Crossings without a site symbol are not additional sites. All nodes in the right-hand diagrams are Fe sites; solid $t_2$ and dashed $t_3$ lines denote effective NNN Fe--Fe hoppings and interchange between A and B for each fixed orbital. The diagonal directions are $\mathbf a_1=a(1,-1)$ and $\mathbf a_2=a(1,1)$. Colors denote basis labels, not lobe signs or orbital order.
(b) Normal bands and (c) Fermi surface evaluated from the uniform two-Fe Bloch Hamiltonian. The gap and chemical potential are obtained from the independent uniform momentum-space self-consistency calculation specified in Appendix~\ref{app:source}.
The Bloch Hamiltonian uses the two-Fe translations and orbital basis defined in Sec.~\ref{sec:method}. The plotted momenta are folded two-Fe coordinates: the horizontal coordinate equals $q_2$ and the vertical coordinate equals $q_1$, with $q_\ell=\mathbf k\cdot\mathbf a_\ell$. The band path connects $\Gamma=(0,0)$, $X=(\pi,0)$ and $M=(\pi,\pi)$ in this folded coordinate system, rather than the Fe-plane momenta used in the QPI maps.}
\end{figure*}

\section{Model}
\label{sec:method}

\subsection{Sample model and observables}

We describe the sample with two Fe sublattices and the active $d_{xz}$ and $d_{yz}$ orbitals. Symmetry-based two-orbital descriptions of Fe--As layers are discussed in Ref.~\cite{HuHao2012S4}; the specific hopping model used here follows Refs.~\cite{Zhang2009Model,Tai2013,Chen2013Disorder}. Figure~\ref{fig:model}(a) illustrates the B-sublattice orbital-label exchange discussed in Appendix A of Ref.~\cite{Chen2013Disorder}; the orbital associated with each color on A and B is stated directly in the legend. This representation changes neither the hopping Hamiltonian nor the numerical states. Figure~\ref{fig:model}(a) depicts the idealized Fe--As coordination environment with $D_{2d}$ symmetry. The model retains only the Fe $d_{xz}$ and $d_{yz}$ orbitals explicitly; the As environment enters through effective Fe--Fe hopping amplitudes. We set the
energy scale by $t=|t_1|=1$; unless stated otherwise, numerical
energies are given in this unit.  The hopping
parameters are $(t_1,t_2,t_3,t_4,t_5,t_6)=
(-1.00,0.08,1.35,-0.12,0.09,0.25)$. Figure~\ref{fig:model} and
Appendix~\ref{app:definitions} specify the fixed two-Fe hopping and orbital convention.

The filling is fixed at $n=1.75$ electrons per Fe and $\mu$ is adjusted during
self-consistency to maintain this filling; the converged values are near
$\mu=-1.324$.  The only pairing interaction is the next-nearest-neighbor (NNN) intraorbital singlet
channel, $J_2=1.2$. We set the onsite Hubbard and Hund couplings to
$U=J_H=0$ and treat $J_2$ as an independent effective pairing interaction.
The normal Hamiltonian therefore contains no onsite Hartree or magnetic
mean field. We use $T=10^{-4}$, with no imposed orbital splitting or hopping strain.  The model is motivated by the hole-doped 12442 iron
arsenide KCa$_2$Fe$_4$As$_4$F$_2$ and related tunneling, angle-resolved photoemission spectroscopy, and vortex work
\cite{Wang2016KCaDiscovery,Wang2016KCaSelfDoping,Wu2020KCaARPES,
Duan2021KCaSTM,Chen2021KCaVortex,Hou2025KCaVortex}. These parameters define an effective two-orbital description. For
physical-length estimates we use the
nearest-neighbor Fe--Fe spacing $a=0.274\,\mathrm{nm}$, obtained from the
reported tetragonal in-plane lattice constant of KCa$_2$Fe$_4$As$_4$F$_2$
\cite{Wang2016KCaDiscovery}.

In real space,
\[
h_{ij}^{\alpha\beta}
=t_{ij}^{\alpha\beta}U_{ij}
+(V_i-\mu)\delta_{ij}\delta_{\alpha\beta},
\qquad U_{ij}=e^{i\varphi_{ij}}.
\]
{\raggedright The real-space BdG Hamiltonian in the reduced singlet Nambu basis
\mbox{$\Psi=(\{c_{i\alpha\uparrow}\},\{c^\dagger_{i\alpha\downarrow}\})^T$} is\par}
\[
\mathcal H=
\begin{pmatrix}h&\Delta\\ \Delta^\dagger&-h^T\end{pmatrix},
\qquad
\Delta_{ij}^{\alpha\beta}=\Delta_{ij}^{\alpha}\delta_{\alpha\beta}.
\]
There are $4N$ states for $N$ Fe sites. At supercell Bloch momentum
$\bm\kappa$ the hole block is $-h^T(-\bm\kappa)$, with the matching
boundary convention in the pairing block. All eigenstate sums below
include both energy signs in this reduced block.
The NNN bond fields satisfy
\[
\Delta_{ij}^{\alpha}
=\frac{J_2}{4}\sum_n
\left(u_{i\alpha n}v_{j\alpha n}^*+u_{j\alpha n}v_{i\alpha n}^*\right)
\tanh\!\left(\frac{E_n}{2T}\right),
\]
with $J_2=0$ away from NNN bonds. The factor $1/4$ is for the full
$4N$-state sum. The local density is
$n_i=\sum_{\alpha n}[|u_{i\alpha n}|^2f(E_n)
+|v_{i\alpha n}|^2(1-f(E_n))]$, with the Fermi function at
$T=10^{-4}$. The fields and chemical potential are determined self-consistently at
$\bm\kappa=0$ on the finite cell, with each state maintaining the prescribed
filling. The spectra and contact responses in Figs.~\ref{fig:scalarJosephson}--\ref{fig:vortexJosephson} are evaluated by supercell Bloch integration with these real-space self-consistent fields held fixed. Figure~\ref{fig:model}(b,c) instead uses an independently obtained uniform superconducting state. The self-consistent chemical potentials and residuals are given in Appendix~\ref{app:source}. QPI is the Fourier transform of the real-space LDOS modulation, not a separate momentum-space self-consistency calculation.

Unless stated otherwise, real-space coordinate pairs throughout the text and captions denote $(x/a,y/a)$, with integer coordinates at Fe sites and half-integer coordinates at plaquette centers. The periodic real-space cell contains $64\times32$ Fe sites and represents a rectangular periodic array of impurities. A nonmagnetic impurity is at
$i_0=(31,15)$ with $V_i=V_{\rm imp}\delta_{i,i_0}$ and
$V_{\rm imp}=0.02,0.05,0.10$.  The two-impurity control has equal potentials
$0.02$ at $(16,8)$ and $(48,24)$.  The vortex state contains no nonmagnetic impurity.
The perpendicular orbital field enters through
\[
\varphi_{ij}=\frac{e}{\hbar}\int_{\mathbf r_i}^{\mathbf r_j}
\mathbf A\cdot d\bm\ell .
\]
The magnetic flux through each one-Fe plaquette is $1/2048$ in units of $h/e$, so the
cell contains two flux quanta $h/2e$.  The vortex-core positions are marked at the plaquette centers $(x/a,y/a)=(15.5,7.5)$ and $(48.5,24.5)$. These half-integer coordinates denote the vortex-core plaquette centers, not Fe-atom positions. Fe atoms occupy integer-coordinate sites; $C_1=(16,8)$ and $N_1=(14,6)$ are nearby Fe sampling sites, not the core centers. With the above Fe--Fe spacing this vortex density corresponds to about
$27\,\mathrm T$. The tip is represented by an ideal $s$-wave reservoir;
Zeeman and spin--orbit couplings are omitted.

The local mean bond amplitude is
\[
A_i=\frac1{16}\sum_{\alpha,\bm\delta}
\left(|\Delta_{i,i+\bm\delta}^{\alpha,\uparrow\downarrow}|
+|\Delta_{i,i+\bm\delta}^{\alpha,\downarrow\uparrow}|\right),
\]
where $\alpha=xz,yz$ and $\bm\delta=\pm a\hat x\pm a\hat y$.  The relative change in the mean bond amplitude is
\[
C_{A,i}=\frac{A_i-A_i^{(0)}}{A_i^{(0)}}.
\]
Here $A_i^{(0)}$ is the mean bond amplitude of the independently self-consistent clean state at the corresponding Fe site. For the condensate phase we use the gauge-covariant local pair field
\[
\psi_i=\frac18\sum_{\alpha,\bm\delta}
U_{i,i+\bm\delta}\Delta_{i,i+\bm\delta}^{\alpha,\uparrow\downarrow}.
\]
The phase of $\psi_i$ defines the condensate winding with the loop convention
in Appendix~\ref{app:phase}; $A_i$ measures the mean bond magnitude.
The local anomalous Green function defined below describes pair propagation at the contact and can be nonzero even though the onsite pairing interaction vanishes.

The spin-summed orbital LDOS is
\begin{align*}
\rho_{i\alpha}(E)=\frac1{N_k}\sum_{\bm\kappa,n}\Big[&
|u_{i\alpha n}^{\bm\kappa}|^2 L_{\eta_s}(E-E_{n\bm\kappa})\\
&+|v_{i\alpha n}^{\bm\kappa}|^2 L_{\eta_s}(E+E_{n\bm\kappa})\Big],
\end{align*}
with $L_{\eta_s}(x)=\eta_s/[\pi(x^2+\eta_s^2)]$ and $\eta_s=0.003$.
The full-spectrum sum includes both spins. The contact-weighted spectrum is
$\rho_i^W(E)=[\rho_{i,xz}(E)+\rho_{i,yz}(E)]/2$, and
$\delta\rho_i^W(E)=\rho_i^W(E)-\rho_{i,0}^W(E)$, where $\rho_{i,0}^W(E)$ is the clean LDOS on the corresponding Fe sublattice. Self-consistent chemical potentials and the LDOS/QPI spectral parameters are listed in Appendix~\ref{app:qpi} (see Fig.~\ref{fig:A1} in Appendix~\ref{app:qpi} for additional real-space spectra).

The principal coherence peaks in the clean LDOS occur at $E_{\rm cp}^{-}=-0.196$ and $E_{\rm cp}^{+}=+0.194$ on the sampled energy grid. We use the positive-energy peak for the main QPI comparison and the negative-energy peak as an additional check. These peak positions are determined at the stated broadening. The QPI amplitude is
\[
Q(\mathbf q,E)=\frac1N\sum_i e^{-i\mathbf q\cdot\mathbf r_i}
\delta\rho_i^W(E).
\]

The maps show $|Q|$ with the same $1/N$ normalization. The rectangular
cell has momentum spacings $\Delta q_x=2\pi/(64a)$ and
$\Delta q_y=2\pi/(32a)$. The momentum sampling and periodic impurity arrangement are therefore direction dependent.

Here $E$ is the sample quasiparticle energy. The quasiparticle tunneling conductance measured with a superconducting tip also depends on the tip density of states \cite{Choubey2024}.

\subsection{Local Josephson response}
\label{sec:jstm_method}

The spatial maps show the local Josephson coupling that determines the current in the weak-tunneling limit.
The tip is a conventional $s$-wave reservoir with gap $\Delta_t=0.20$, unless otherwise stated.  Two independent orbital-diagonal
contact channels are used with $w_{xz}=w_{yz}=1/2$.  A coherent contact to a single orbital combination would also contain
interorbital anomalous terms.
The local sample Green matrix is
$\bar G_{s,i}(z)=N_k^{-1}\sum_{\bm\kappa}
[(z-\mathcal H(\bm\kappa))^{-1}]_{ii}$ in the contact order
$(e_{xz},e_{yz},h_{xz},h_{yz})$. Its interorbital elements are retained.
All contact quantities use the same sample phase convention and tip--sample phase reference; phases are not independently reset at different sites. To fix the sign
relative to the tunneling Hamiltonian below, define
\begin{equation}
 F_{i\alpha}(z)=-[\bar G_{s,i}(z)]_{e_\alpha h_\alpha}
 =-\frac1{N_k}\sum_{\bm\kappa n}
 \frac{u_{i\alpha n}^{\bm\kappa}v_{i\alpha n}^{\bm\kappa *}}
 {z-E_{n\bm\kappa}}.
 \label{eq:Fconvention}
\end{equation}
The eigenvectors and anomalous residues use this fixed contact phase convention. With $F^{(+)}(i\omega)=[F(i\omega)+F(-i\omega)]/2$,
\begin{align}
 K_{i\alpha}&=\int_0^\infty\frac{d\omega}{\pi}
 \frac{\Delta_t F_{i\alpha}^{(+)}(i\omega)}
 {\sqrt{\omega^2+\Delta_t^2}},\\
 K_i&=\sum_{\alpha=xz,yz}w_\alpha K_{i\alpha}
 \equiv K_{R,i}+iK_{I,i}.
 \label{eq:K}
\end{align}
We evaluate the Josephson coefficient in the $T\to0$ limit using the self-consistent sample fields and integrating over the full positive imaginary-frequency axis.

For tunneling strength $\Gamma_c$, the weak current is
\begin{equation}
 \bar I_{w,i}(\varphi)=2\Gamma_c\operatorname{Im}(e^{i\varphi}K_i),
 \qquad I_{w,i}=\frac{2et}{\hbar}\bar I_{w,i}.
 \label{eq:Iweak}
\end{equation}

The clean coefficient is $K_{\rm clean}\simeq6.47\times10^{-5}>0$ at
$\Delta_t=0.20$. For a time-reversal-preserving
sample and nonmagnetic contact, $K_i$ is real in this convention.
At other tip gaps clean $K$ may be negative; a reversal relative to clean
means opposite signs at the same $\Delta_t$ using the same tip--sample phase reference.
Determining the sign of the Josephson coupling experimentally requires phase-sensitive information \cite{Karan2022}.

Maps for the vortex states are calculated in a fixed gauge with a common phase reference and charge-$2e$ boundary conditions.
The local Josephson coupling depends on the tip position and the orbital tunneling weights
\cite{Smakov2001,Choubey2024}.

\label{sec:controls}

The pair field is determined self-consistently for both the clean and impurity states at the same filling. For the spectral analysis we retain the BdG states satisfying
$|E_{n\bm\kappa}|\le E_c$ in the spectral sum and write
\begin{equation}
\delta K_i^{(\le E_c)}=
K_i^{(\le E_c)}-K_{\rm clean}^{(\le E_c)}.
\label{eq:increment}
\end{equation}
The difference and the individual cumulative sums distinguish the
impurity-induced redistribution from the background of the clean contact.
The cumulative coupling contains contributions from sample states with $|E_{n\bm\kappa}|\le E_c$, each with its full Josephson tunneling weight. The local spectral comparisons use the impurity center
$(31,15)$ and the off-center site $(29,13)$. Other nearby sites are identified
by their coordinates.

\subsection{Finite tunneling strength}
\label{sec:contactcheck}

We retain the local tunneling Hamiltonian to
all orders between superconducting electrodes with fixed pair potentials \cite{Cuevas1996}.
The total tunneling strength is $\Gamma_c$, with
$\Gamma_\alpha=w_\alpha\Gamma_c$ for the two independent tip channels.
It is distinct from both spectral broadening and transmission.

Let $\Sigma_t$ be the tip self-energy in the local orbital--Nambu basis and
$M_i=1-\bar G_{s,i}\Sigma_t$. The tip Green function and tunneling matrix
are specified in Appendix~\ref{app:frequency}.
The full current is
\begin{equation}
 \bar I_i(\varphi)=\int_0^\infty\frac{d\omega}{\pi}
 \operatorname{Re}\operatorname{Tr}
 [M_i^{-1}\bar G_{s,i}\partial_\varphi\Sigma_t],
 \qquad I_i=\frac{2et}{\hbar}\bar I_i.
 \label{eq:Ifull}
\end{equation}
Its order-$\Gamma_c$ expansion yields Eq.~\eqref{eq:Iweak}. The sample
Bloch average is taken before the local matrix inverse.

The normal conductance is evaluated with the sample pair potential set to zero, while retaining the normal Hamiltonian and chemical potential of the corresponding superconducting state.
With $\nu_{N,i}=-\pi^{-1}\operatorname{Im}\operatorname{Tr}
[w g_{N,i}(i\eta_N)]$ the weighted normal density of states (DOS) per spin, the weak
conductance is $G_{N,w}=G_0\,4\pi\Gamma_c\nu_{N,i}$,
where $G_0=2e^2/h$ includes spin degeneracy. For $R_{N,w}=G_{N,w}^{-1}$,
\begin{equation}
 \frac{e I_{c,w}R_{N,w}}{\Delta_t^{\rm phys}}
 =\frac{|K_i|}{\Delta_t\nu_{N,i}},\qquad
 \Delta_t^{\rm phys}=t\Delta_t.
\end{equation}
For comparison, the equal-gap tunnel limit gives the Ambegaokar--Baratoff
result \cite{Ambegaokar1963,Ota2010Multigap}.

Normal conductance for the clean and nonmagnetic-impurity samples uses broadening $\eta_N=0.00075$ in the normal-state Green function. The resulting normal conductance sets the weak-tunneling current--resistance scale; its sensitivity to broadening and Bloch integration is reported in Appendix~\ref{app:frequency}.
Vortex CPRs are presented with weak-tunneling normalization.

We use the Fourier convention
\begin{equation}
 I_i(\varphi)=\sum_{n\ge1}
 [a_{n,i}\sin(n\varphi)+b_{n,i}\cos(n\varphi)].
 \label{eq:harmonics}
\end{equation}
The $m=1$ Josephson harmonic has the signed sine coefficient $a_1$. All normalized CPRs and harmonics use
the positive weak-tunneling amplitude at the same state, site and coupling,
\[
 I_{c,w}(i,\Gamma_c)=\frac{2et}{\hbar}\,2\Gamma_c|K_i|.
\]
Finite transmission affects Josephson transport in superconducting atomic contacts. Della Rocca et al. measured the current--phase relation \cite{DellaRocca2007}, while Senkpiel et al. observed deviations from the Ambegaokar--Baratoff relation in a high-transmission single-channel contact \cite{Senkpiel2020}.

\begin{figure*}[!t]
\centering
\includegraphics[width=450bp]{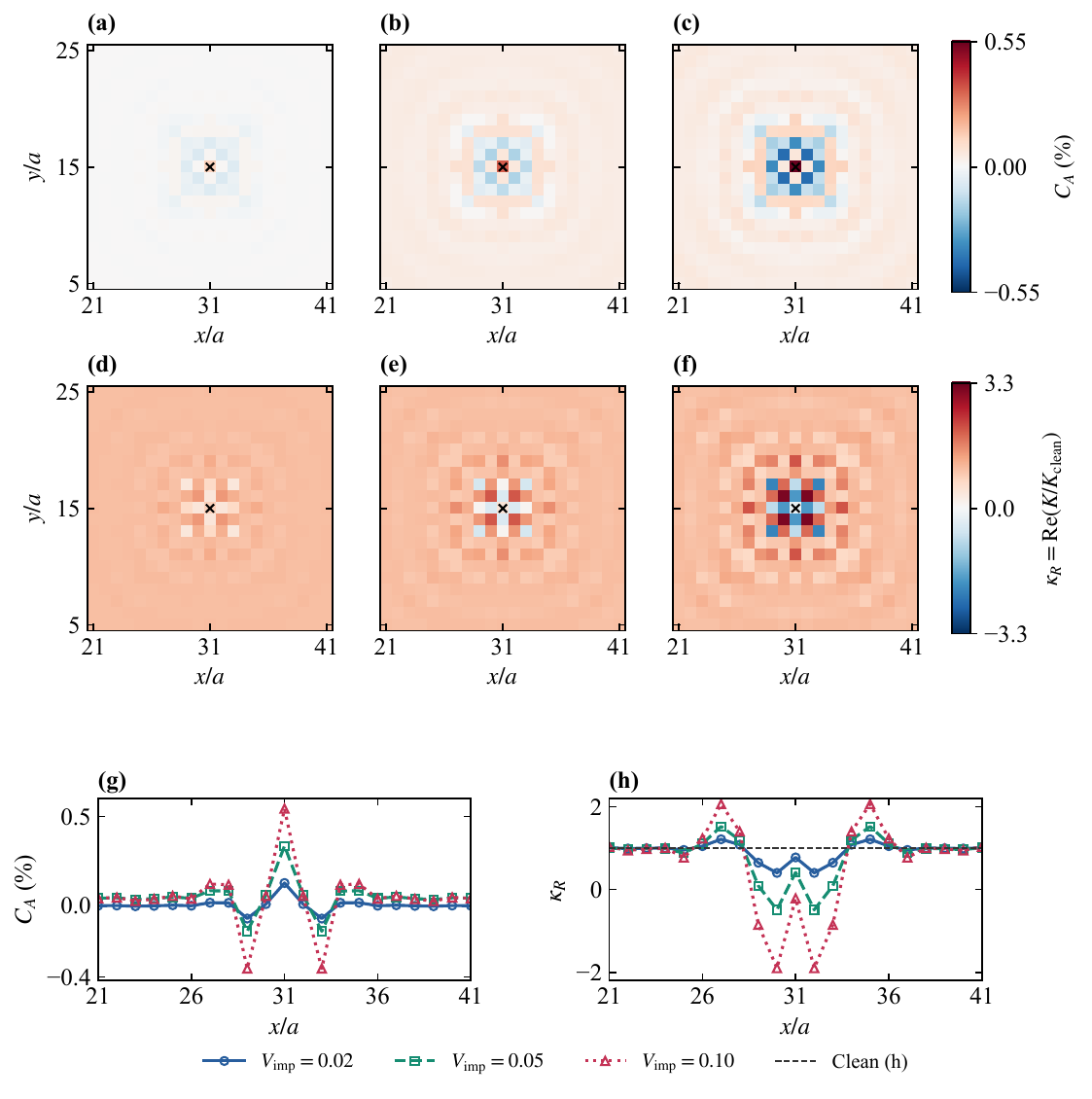}
\caption{\label{fig:scalarJosephson}Response of real-space self-consistent states to nonmagnetic impurities at
$\Delta_t=0.20$. Panels (a,d), (b,e) and (c,f) use $V_{\rm imp}=0.02,0.05,0.10$, respectively. (a--c) Relative change $C_{A,i}$ in the mean bond amplitude, displayed in percent;
(d--f) $\kappa_{R,i}=\operatorname{Re}(K_i/K_{\rm clean})$ in the common phase
reference; (g,h) cuts along $y/a=15$. The normalized clean coupling in (h) is unity.}
\end{figure*}

\section{Nonmagnetic Impurity}
\label{sec:josephsonresults}

\label{sec:scalarjosephson}

The local Josephson coupling changes with nonmagnetic impurity potential at $\Delta_t=0.20$ (Fig.~\ref{fig:scalarJosephson}). At the impurity center,
$K/K_{\rm clean}$ is approximately $0.777$, $0.413$ and $-0.211$ for
$V_{\rm imp}=0.02$, $0.05$ and $0.10$. The corresponding numbers of sites with negative Josephson coupling in the full cell are $0$, $8$ and $13$. Off-center negative couplings are present at $V_{\rm imp}=0.05$, whereas the impurity-center coupling becomes negative only at $V_{\rm imp}=0.10$ among the sampled potentials.

The NNN bond pair fields show no $\pi$ phase shift relative to their clean counterparts at the three impurity strengths. Individual bond amplitudes change by up to about $2\%$ for $V_{\rm imp}=0.10$. The
Josephson coefficient nevertheless changes sign without a corresponding sign
reversal of the bond pair field.

Positive and negative anomalous contributions at the impurity center are each of order $10^{-2}$, while their net Josephson coupling is of order $10^{-5}$ (see Fig.~\ref{fig:A14} in Appendix~\ref{app:frequency}). This near cancellation makes the coupling sensitive to a small impurity-induced spectral redistribution, allowing it to reverse sign while the bond pair fields change only weakly.

\begin{figure*}[!t]
\centering
\includegraphics[width=400bp]{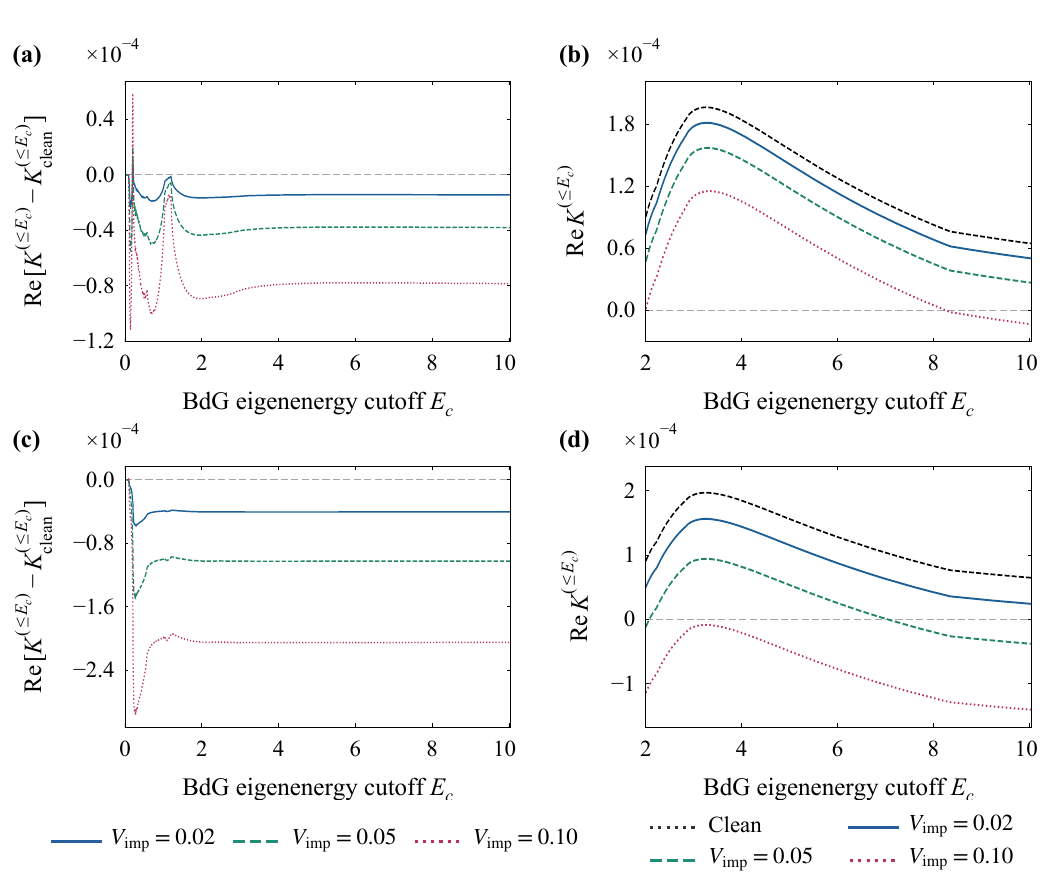}
\caption{\label{fig:spectralComp}Energy-resolved contributions from the real-space BdG sample states to the local Josephson coupling at $\Delta_t=0.20$. Panels (a,c): impurity-induced cumulative difference
$\operatorname{Re}\delta K_i^{(\le E_c)}$. Panels (b,d): individual cumulative coefficients for $E_c\ge2$,
retaining all contributions below 2. Panels (a,b) refer to the center $(31,15)$; panels (c,d) refer to the off-center site $(29,13)$. $E_c$ is a BdG
eigenenergy cutoff, not a bias voltage or an imaginary-frequency cutoff.}
\end{figure*}

\section{Spectral Decomposition}
\label{sec:spectralcomp}

The impurity-induced spectral change and the cumulative Josephson coupling resolve the redistribution of anomalous contributions and their net balance (Fig.~\ref{fig:spectralComp}). The difference
$\delta K_i^{(\le E_c)}$ approaches its final value before the individual clean
and impurity coefficients do. At larger $E_c$, the individual clean and impurity coefficients continue to change in a similar way. The impurity changes the lower-energy contributions, while a broader range of states contributes similarly to the clean and impurity couplings. The sign of the total coupling depends on the balance of contributions over the full spectrum.

At the impurity site for $V_{\rm imp}=0.10$, higher-energy states are essential to the negative sign of the total Josephson coupling, although the impurity-induced change is concentrated at lower energies. The cumulative coefficient is positive at intermediate $E_c$ and becomes negative only after additional spectral weight is included. States at energies exceeding both the coherence-peak energy of the clean sample and the tip gap contribute to the total Josephson coupling.  The decomposition
resolves contributions by BdG eigenenergy; assigning them to normal-state
Fermi-surface sheets requires an additional band projection. Cancellation between Josephson contributions is also possible in multigap and $s$--$s_{\pm}$ junctions
\cite{Agterberg2002Multigap,Ota2009Multigap,Sperstad2009Transport,
Koshelev2012}. The nonmagnetic impurity changes this balance in a position-dependent way.

\begin{figure*}[!t]
\centering
\includegraphics[width=400bp]{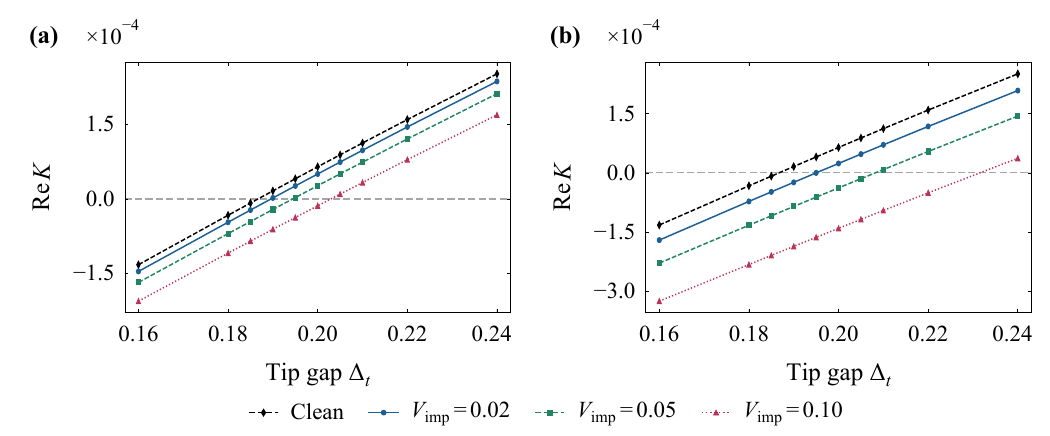}
\caption{\label{fig:tipGap}Tip-gap dependence of the signed Josephson
coefficient at (a) the impurity center $(31,15)$ and (b) the off-center site $(29,13)$. The sample
states and global phase convention are unchanged throughout the scan. Lines connect the calculated coefficients at the sampled tip gaps; the zero crossings are not fitted.}
\end{figure*}

\section{Tip Dependence}
\label{sec:contactresults}

\subsection{Tip gap}
\label{sec:tipgap}

Changing the $s$-wave tip gap alters the energy weighting of the pair-tunneling contributions without changing the self-consistent sample spectrum (Fig.~\ref{fig:tipGap}). The clean coefficient passes through zero between the sampled gaps $0.185$ and $0.190$, while the impurity shifts the zero in a position-dependent way. For $V_{\rm imp}=0.10$, the coupling at the impurity center is negative at $\Delta_t=0.20$ and positive at $0.205$. The zero crossing at the off-center site occurs at a larger tip gap than at the impurity center. The impurity-induced reversal relative to the clean contact persists over finite intervals of the tip gap.

\begin{figure*}[!t]
\centering
\includegraphics[width=450bp]{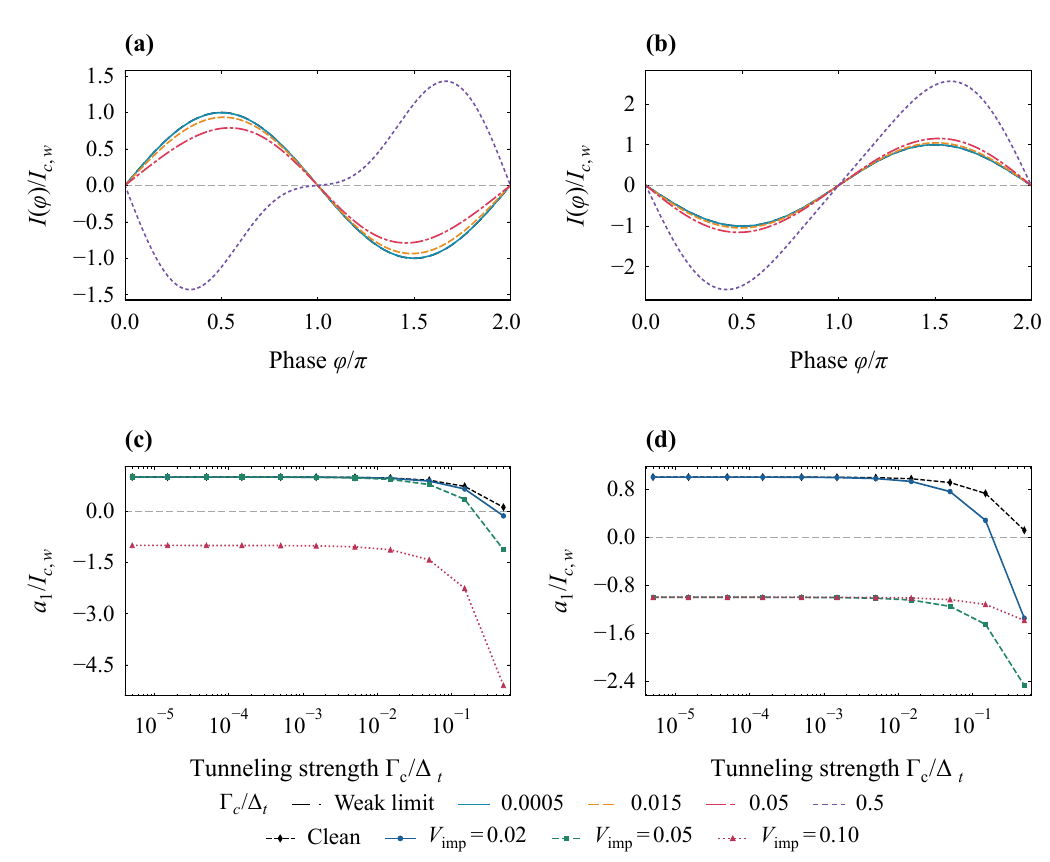}
\caption{\label{fig:finiteContact}Effect of tunneling strength on the Josephson current for the same self-consistent sample states. Panels (a,c) refer to the impurity center $(31,15)$ and panels (b,d) to the off-center site $(29,13)$.
Upper row: CPRs for $V_{\rm imp}=0.05$. Lower row: signed sine coefficient $a_1$ of the $m=1$ Josephson harmonic for the clean and nonmagnetic-impurity contacts. Each curve is normalized by the positive weak-tunneling amplitude for the same sample state, contact site and tunneling strength. In (c,d), the tunneling-strength axis is logarithmic.}
\end{figure*}

\subsection{Tunneling strength}
\label{sec:finiteharmonic}

The CPR changes with tunneling strength while each self-consistent sample state remains fixed (Fig.~\ref{fig:finiteContact}). For $V_{\rm imp}=0.05$, the weak-tunneling sine coefficient $a_1$ of the $m=1$ Josephson harmonic is positive at the impurity center and negative at the off-center site. At the largest displayed $\Gamma_c/\Delta_t$, $a_1$ at the center has also become negative.  This contact-induced sign reversal of $a_1$ occurs at both sampled sites for $V_{\rm imp}=0.02$ and at the center for $V_{\rm imp}=0.05$. The off-center coefficient for $V_{\rm imp}=0.05$ and both coefficients for $V_{\rm imp}=0.10$ are already negative in the weak limit. For the clean contact, $a_1$ remains positive over the displayed range while the CPR develops higher harmonics.

Changing the tunneling strength can therefore reverse the sign of $a_1$ without changing the sample pair field. The full CPR also contains higher harmonics, so
the sign of $a_1$ alone does not determine the equilibrium phase of the complete
junction. Corrections beyond the weak-tunneling limit can change the magnitude and sign of the $m=1$ sine coefficient when anomalous spectral contributions nearly cancel \cite{Sperstad2009Transport,Koshelev2012}.

For the zero-field contacts, the small weak-tunneling current--resistance product reflects cancellation of anomalous pair-tunneling contributions, whereas the weak-tunneling normal conductance is proportional to the contact-weighted normal-state density of states. The same-contact normal resistance gives a weak-tunneling
$eI_{c,w}R_{N,w}/\Delta_t^{\rm phys}$ of order $10^{-3}$ for the clean contact, with a
smaller value at the impurity center for $V_{\rm imp}=0.10$ and a larger value at the selected
off-center site (see Fig.~\ref{fig:A8} in Appendix~\ref{app:frequency}). The respective values are approximately $1.84\times10^{-3}$,
$3.75\times10^{-4}$ and $4.01\times10^{-3}$ at $\eta_N=0.00075$. At the same tunneling strength, the deviation from the weak-tunneling CPR depends on contact position (see Fig.~\ref{fig:A12} in Appendix~\ref{app:frequency}). 

\begin{figure*}[!t]
\centering
\includegraphics[width=400bp]{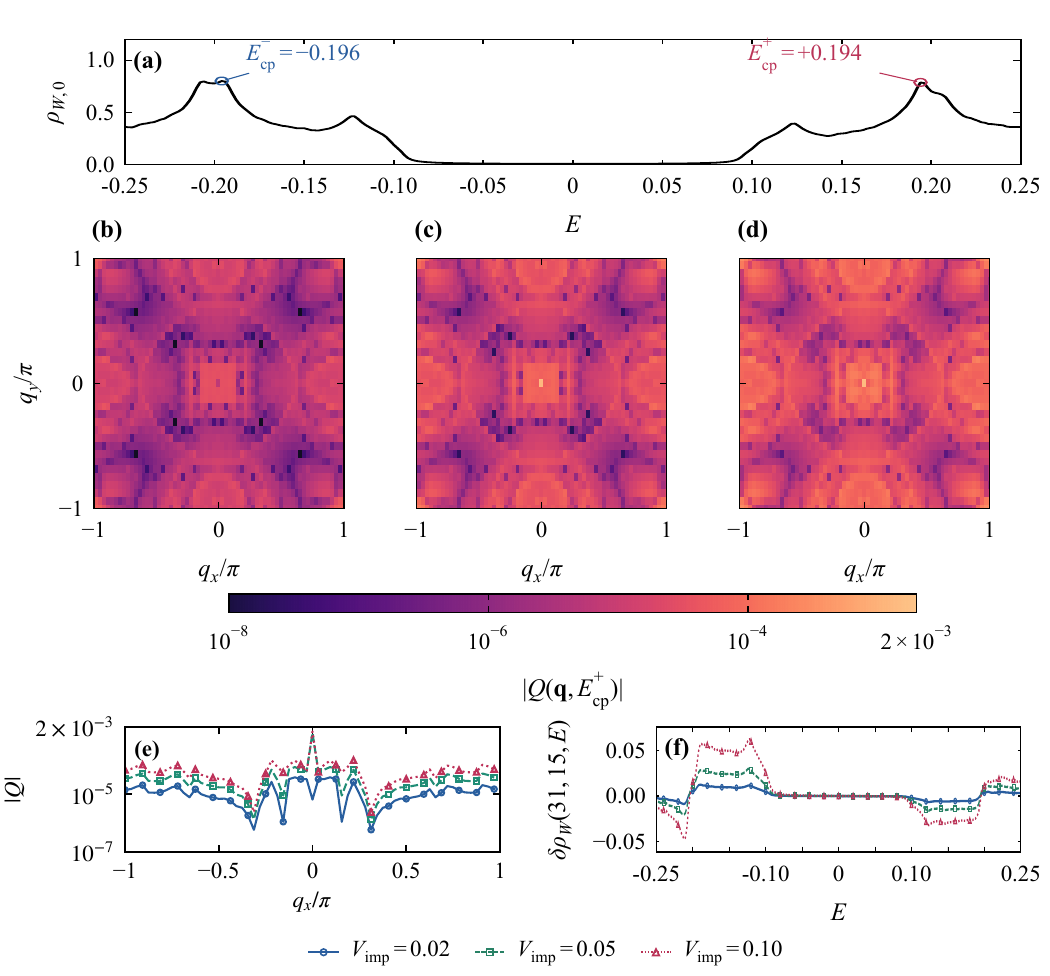}
\caption{\label{fig:scalarQPI}Quasiparticle interference in self-consistent real-space impurity states for different nonmagnetic impurity potentials.
(a) Clean contact-weighted LDOS and its sampled coherence peaks.
(b--d) $|Q(\mathbf q,+0.194)|$ for $V_{\rm imp}=0.02,0.05,0.10$, with a
common absolute color scale.  (e) Full $q_y=0$ cut at $E=+0.194$.  (f) Signed LDOS change
at the impurity center.  The maps include the impurity-induced spatial variation of the self-consistent pair potential on the rectangular periodic cell. Panels (b--d) use a logarithmic color scale, and the vertical axis of the momentum cut in (e) is logarithmic.}
\end{figure*}

\section{QPI}
\label{sec:normal}

\label{sec:scalarqpi}

The QPI is calculated from the self-consistent real-space impurity states, including the impurity-induced spatial variation of the pair potential (Fig.~\ref{fig:scalarQPI}). QPI describes the impurity modulation of the electron LDOS in the superconducting state and depends on superconducting coherence factors. The comparison uses the coherence-peak energy of the clean sample.  At $E_{\rm cp}^{+}=0.194$, the $q_y=0$ cuts have peaks at $q_x/\pi=3/32$ and $7/32$ whose relative QPI amplitudes change with impurity potential.
The nonzero-$q$ maximum amplitudes at this energy are
$7.24\times10^{-5}$, $1.72\times10^{-4}$ and $3.14\times10^{-4}$,
respectively, under the same $1/N$ definition. The corresponding
$|Q(0)|$ values are $6.38\times10^{-6}$, $1.37\times10^{-3}$ and
$1.42\times10^{-3}$; the zero-momentum contribution is reported separately from the finite-momentum structures. The zero-momentum component is the spatially averaged LDOS change, while finite momenta describe its spatial modulation. The dependence of QPI on impurity strength beyond the Born approximation has been discussed in Refs.~\cite{Akbari2013FullT,Hong2021}.

The QPI maps evaluated at the negative-energy clean coherence peak and at the inner clean LDOS maximum $E=0.124$ differ from those at $E=0.194$ (see Fig.~\ref{fig:A6} in Appendix~\ref{app:qpi}). The anisotropy shown here pertains to the self-consistent state in the rectangular periodic cell. For these impurity states, the local Josephson coupling can reverse sign while the relative QPI peak amplitudes change at the selected coherence-peak energy.

\begin{figure*}[!t]
\centering
\includegraphics[width=450bp]{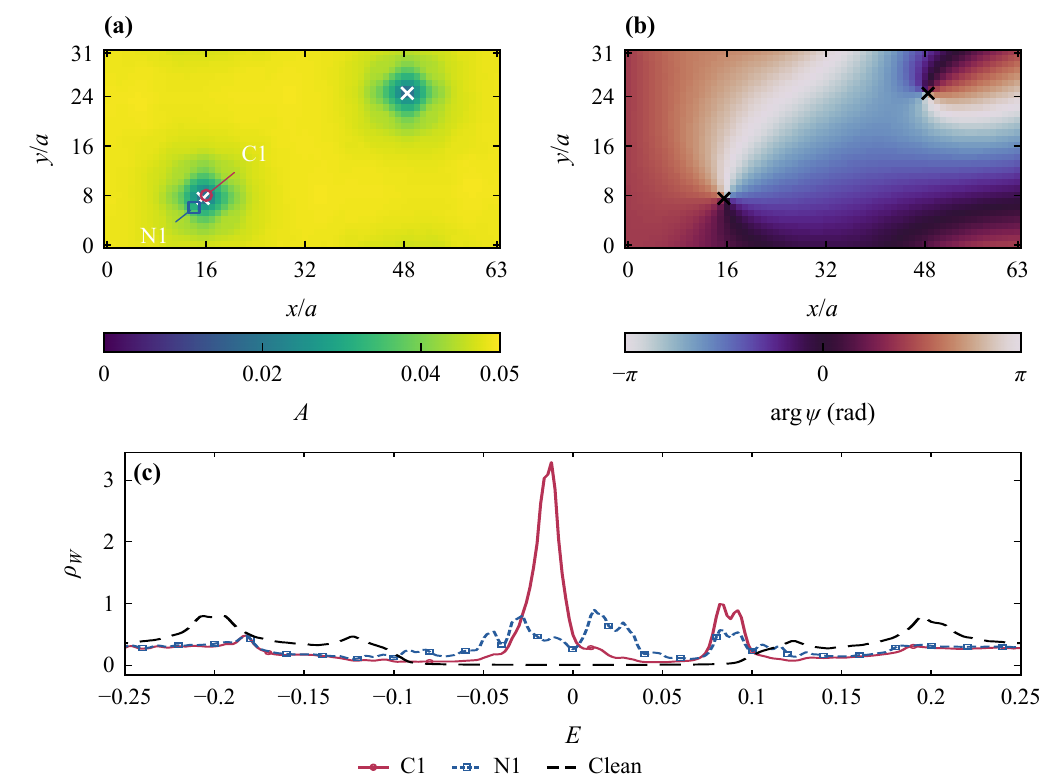}
\caption{\label{fig:vortex}Real-space self-consistent two-vortex sample.
(a) Mean NNN bond magnitude.  (b) Phase of the local bond pair field $\psi_i$; crosses mark the half-integer core plaquette centers.  (c) Contact-weighted LDOS at
$C_1=(16,8)$, $N_1=(14,6)$ and in the clean superconducting state.}
\end{figure*}

The additional spectral comparisons include two-impurity and vortex states. Vortex-induced quasiparticle spectra have been studied in iron-based superconductors \cite{Yin2009Vortex,Wang2010Vortex,Shan2011Vortex,Zhou2011Vortex}.
Additional spatial spectra for one nonmagnetic impurity, two nonmagnetic impurities and the two-vortex state, and QPI maps for the latter two cases, are shown in Fig.~\ref{fig:A7} in Appendix~\ref{app:qpi}.
The two identical impurities are separated by $(\Delta x/a,\Delta y/a)=(32,16)$, a half-cell translation. If the self-consistent LDOS modulation has this periodicity, it produces destructive interference in the QPI at selected wave vectors.  Such a two-center arrangement can change the map without
vortex winding. Repeated scattering between two impurities has also been studied \cite{Ding2023Twin}. In the vortex state, the orbital magnetic field and spatial variations of the pair amplitude and phase all modify the quasiparticle spectrum.

\section{Vortex}
\label{sec:vortices}

\label{sec:vortexsample}

Vortices produce spatial variations in both the amplitude and phase of the superconducting pair field. We compare the magnitude and phase of the local Josephson response with those of the condensate in a self-consistent two-vortex state. The mean NNN bond amplitude is suppressed near the cores, while the phase of the local bond pair field $\psi_i$ winds around each core (Fig.~\ref{fig:vortex}). For counterclockwise loops in the Fe plane, both cores have condensate winding $m_\psi=-1$ in the positive-field state, and reversing the field reverses the winding to $m_\psi=+1$.

At $C_1=(16,8)$, the LDOS has a low-energy peak; at the nearby site $N_1=(14,6)$, the low-energy spectral weight is distributed among several peaks.
Vortex-core LDOS and its spatial variation have been studied in iron-based systems and multiorbital BdG models \cite{Hu2009Vortex,Wang2010Vortex,Shan2011Vortex,Zhou2011Vortex,Gao2011Vortex}.  We compare their Josephson responses in the same self-consistent vortex state.

\label{sec:vortexjosephson}

\begin{figure*}[!t]
\centering
\includegraphics[width=450bp]{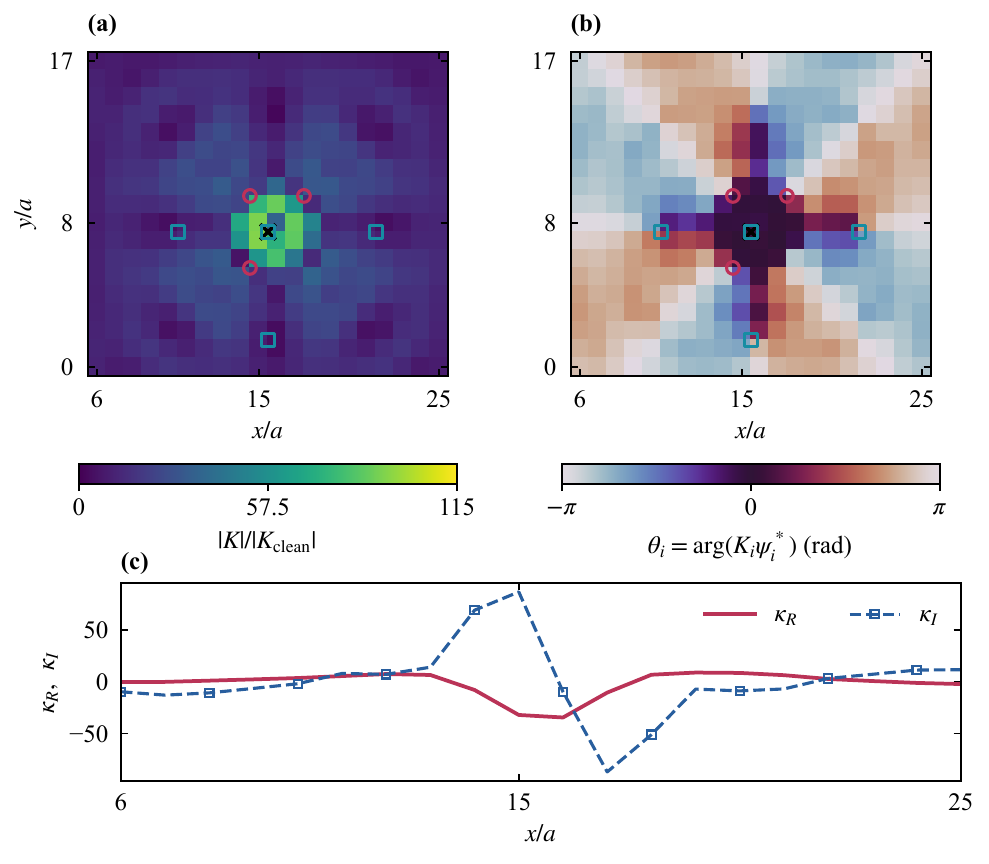}
\caption{\label{fig:vortexJosephson}Local Josephson response near the core plaquette center $(15.5,7.5)$.
(a) Magnitude relative to the clean Josephson coupling for the same $s$-wave tip; (b) phase relative
to the condensate, $\theta_i=\arg(K_i\psi_i^*)$; (c) real and
imaginary components along $y/a=8$.
Circles and squares mark selected plaquettes with Josephson-response windings $m_K=+1$ and $-1$, respectively, as specified in Appendix~\ref{app:phase}. The condensate core is marked separately.}
\end{figure*}

The magnitude of the local Josephson response and its phase relative to the local bond pair field have distinct spatial structures (Fig.~\ref{fig:vortexJosephson}). The gauge-invariant relative phase $\theta_i=\arg(K_i\psi_i^*)$ is the phase difference between the Josephson response and the local bond pair field. This difference varies on the Fe-lattice scale at some positions. The coefficient $K_i$ combines anomalous contributions from different sample energies with the energy-dependent weight of the tip, so the response phase and the bond-pair phase can vary differently in space. The response near the second core and the energy-resolved contributions to the complex Josephson coupling are shown in Figs.~\ref{fig:A4} and \ref{fig:A16} in Appendix~\ref{app:phase}.

The symbols mark plaquettes with Josephson-response winding numbers $m_K=+1$ and $-1$, evaluated with the magnetic-link convention in Appendix~\ref{app:phase}. The color map shows the relative phase $\theta_i$. On the smallest common loops around the two cores, the Josephson response and condensate have the same winding; larger loops enclose additional winding of the Josephson response while the condensate winding remains unchanged (see Table~\ref{tab:loopwinding} in Appendix~\ref{app:phase}). These windings describe the local Josephson response and do not identify additional magnetic-flux vortices.
Vortex-induced pairing components can have spatial phase variations different from those of the dominant condensate
\cite{Berlinsky1995,Franz1996Vortex,Ichioka1996Vortex,Heeb1996Vortex,
XuRenTing1996Vortex}. Superconducting-tip probes of vortex-core pairing symmetry have also been proposed \cite{Yokoyama2008Vortex}. Here we examine how the even-frequency, orbital-diagonal anomalous correlations contribute to the local Josephson response. States obtained from different initial pair fields and vector potentials also contain two vortex cores (see Figs.~\ref{fig:A2} and \ref{fig:A3} in Appendix~\ref{app:phase}).

The ratio $|K_i|/|K_{\rm clean}|$ compares Josephson couplings, with $|K_{\rm clean}|$ suppressed by cancellation between anomalous spectral contributions. It does not directly measure the ratio of bond-pair-field magnitudes. The real and imaginary components of $K_i$ depend on the gauge and phase reference, whereas its phase relative to $\psi_i$ is gauge invariant. The normalized current--phase relations at $C_1$ and $N_1$ and their harmonic content are shown in Figs.~\ref{fig:A10}--\ref{fig:A12} in Appendix~\ref{app:frequency}.

Near core 1, both $A_i$ and $|\psi_i|$ are suppressed and recover outward, whereas $|K_i|$ has direction-dependent peaks and minima (see Fig.~\ref{fig:vortexprofiles} in Appendix~\ref{app:phase}). For each orbital, the pair fields on the two diagonal bond directions already differ in the clean state, and this difference increases in magnitude at some sites near the vortices. These local pairing-field changes do not by themselves explain the additional winding of the Josephson response.

\section{Discussion}
\label{sec:discussion}

Anomalous spectral contributions of opposite sign nearly cancel in the clean contact, leaving a small net Josephson coefficient. The self-consistent impurity state redistributes these contributions locally, so a comparatively small spectral change can reverse the weak-tunneling coupling without a corresponding $\pi$ phase shift of the bond pair fields. The change is concentrated at lower energies, but the sign of the total coupling depends on contributions over a broader energy range. The finite-$U$ comparison shows how charge feedback modifies this balance: suppression of the impurity-center coupling is weaker in the checked nonmagnetic $U=1$ states, while the coefficient at the checked neighboring site remains negative with a smaller magnitude than at $U=0$ (see Fig.~\ref{fig:finiteuSCF} and Table~\ref{tab:finiteu} in Appendix~\ref{app:finiteu}). These results concern the checked nonmagnetic branch and positions.

For some contacts, spatial variations of the Josephson current correlate with the local pair amplitude \cite{Graham2017}. In the disordered models of Ref.~\cite{Sulangi2021}, the Josephson critical current remains correlated with the pair amplitude even when the pair amplitude and spectroscopic gap are weakly correlated. Contact-dependent zeros and sign changes have also been obtained for an $s$-wave tip on a $d$-wave surface \cite{Choubey2024}, and Josephson contributions from different bands can cancel \cite{Sperstad2009Transport,Koshelev2012}. In Josephson tunneling through a magnetic-impurity Yu--Shiba--Rusinov state, a change in current sign need not result from a local pair-field sign change \cite{Theiler2026}. In the present nonmagnetic-impurity states, the position-dependent sign changes arise from the balance of positive and negative anomalous spectral contributions in the self-consistent sample. A local Josephson amplitude or sign therefore cannot be identified directly with the magnitude or sign of the bond pair field.

The tip-gap dependence follows from the energy weighting of the sample anomalous spectrum. Varying the tip gap changes this weighting and can move the coupling through zero while the sample state remains unchanged. Increasing the tunneling strength can also reverse the $m=1$ sine coefficient $a_1$, but higher harmonics contribute to the current, and the equilibrium phase must be obtained from the full Josephson energy. The contact considered here has independent orbital channels with equal weights. Measuring the sign of $a_1$ requires a phase reference \cite{Karan2022}; the nonnegative critical-current amplitude alone does not determine that sign. The same-contact $I_{c,w}R_{N,w}$ values are weak-tunneling current--resistance products and retain the broadening and Bloch-integration dependence of the normal resistance. The calculation at finite tunneling strength describes equilibrium transport between fixed superconducting reservoirs \cite{Cuevas1996}.

QPI provides a spectral comparison with the Josephson response: at the selected coherence-peak energy, increasing impurity strength mainly changes the relative intensities of existing QPI features, whereas the energy-integrated anomalous contributions can change the sign of the local Josephson coupling. In the self-consistent two-vortex state, the comparison extends to the response phase relative to the condensate. The response and condensate share the winding on small loops around each core, while larger loops enclose additional winding of the Josephson response. The pair-field magnitude is suppressed near the cores, whereas $|K|$ has direction-dependent peaks and minima. These observations establish a spatially varying relative phase on the sampled lattice; the local pairing-field changes alone do not determine the spectral origin of the additional response winding.

\section{Conclusion}
\label{sec:conclusion}

Nonmagnetic impurities can reverse the local Josephson coupling without a corresponding $\pi$ phase shift of the bond pair fields. For the local $s$-wave tip considered here, the clean coupling is small because anomalous contributions of opposite sign nearly cancel. Their redistribution in the self-consistent impurity state produces negative coupling away from the impurity at a weaker potential than at the impurity site. The impurity-induced change is concentrated at lower energies, while states over a broader energy range contribute to the total clean and impurity couplings. The coupling sign also depends on the tip gap, and stronger tunneling can reverse the $m=1$ sine component of the current for an unchanged sample state. At the selected coherence-peak energy, increasing impurity strength mainly changes the relative intensities of existing QPI features.

In the self-consistent two-vortex state, the local Josephson phase varies spatially relative to the condensate. The response and condensate share the winding close to each core, while larger loops enclose additional winding of the Josephson response. The spatial variation of $|K|$ also differs from the suppression and recovery of the pair-field magnitude. In both impurity and vortex states, the local Josephson response depends on the anomalous spectrum and the tip--sample coupling. Its spatial variation and phase need not coincide with those of the bond pair field.

\section*{Acknowledgements}
We thank L. Shan, Y. Zong and J.-X. Yin for helpful discussions.
Y. Z. acknowledges support from the National Natural Science Foundation of China (Grant Nos. 12474153 and 11804163) and the Guangdong Provincial Quantum Science Strategic Initiative (GDZX2401001).

\appendix
\setcounter{dbltopnumber}{4}
\renewcommand{\dbltopfraction}{0.90}
\flushbottom
\setlength{\textfloatsep}{10pt}
\setlength{\intextsep}{10pt}
\setlength{\dbltextfloatsep}{10pt}
\setlength{\floatsep}{10pt}
\setlength{\dblfloatsep}{10pt}
\predisplaypenalty=0
\setlength{\abovedisplayskip}{10pt}
\setlength{\belowdisplayskip}{10pt}
\setlength{\abovedisplayshortskip}{6pt}
\setlength{\belowdisplayshortskip}{6pt}
\makeatletter
\setlength{\@fptop}{0pt}
\setlength{\@dblfptop}{0pt}
\setlength{\@fpsep}{10pt}
\setlength{\@dblfpsep}{10pt}
\setlength{\@fpbot}{0pt plus 1fil}
\setlength{\@dblfpbot}{0pt plus 1fil}
\makeatother
\section{Model and numerical conditions}
\label{app:source}
\label{app:definitions}
\label{app:dyson}
The hopping parameters are $t_1=-1.00$, $t_2=0.08$, $t_3=1.35$, $t_4=-0.12$, $t_5=0.09$,
and $t_6=0.25$.  Here $t_1$ and $t_5$ are nearest-neighbor intraorbital and
interorbital hoppings, $t_{2,3}$ are staggered next-nearest-neighbor (NNN)
intraorbital hoppings, $t_4$ is NNN interorbital hopping, and $t_6$ is intraorbital hopping over $\pm2a\hat x$ and
$\pm2a\hat y$.  The symbols below refer to the displayed two-Fe matrix.
A sublattice-dependent orbital relabeling transforms the basis and the
Hamiltonian matrix together. We keep the
displayed basis fixed throughout the calculation.

Using two-Fe translations $\mathbf a_1=a(1,-1)$ and
$\mathbf a_2=a(1,1)$, with $q_\ell=\mathbf k\cdot\mathbf a_\ell$, the
zero-field normal Hamiltonian in the fixed-orbital basis
$\Psi_{\mathbf q}=(c_{xz,A},c_{yz,A},c_{xz,B},c_{yz,B})^T$ is
\[
h_0(\mathbf q)=
\begin{pmatrix}
E_H&E_{12}&E_t&E_c\\
E_{12}&E_V&E_c&E_t\\
E_t&E_c&E_V&E_{12}\\
E_c&E_t&E_{12}&E_H
\end{pmatrix}-\mu\mathbb I_4,
\]
where
\begin{align*}
E_H&=2t_2\cos q_1+2t_3\cos q_2+4t_6\cos q_1\cos q_2,\\
E_V&=2t_3\cos q_1+2t_2\cos q_2+4t_6\cos q_1\cos q_2,\\
E_{12}&=2t_4(\cos q_1+\cos q_2),\\
E_t&=4t_1\cos(q_1/2)\cos(q_2/2),\\
E_c&=4t_5\cos(q_1/2)\cos(q_2/2).
\end{align*}
The $t_2/t_3$ assignments interchange between the two Fe sublattices as shown
in Fig.~\ref{fig:model}.
The normal bands and Fermi surface in Fig.~\ref{fig:model}(b,c) follow from the uniform two-Fe Bloch Hamiltonian; an independently obtained uniform superconducting state determines the gap and chemical potential at $n=1.75$, with $\mu\simeq-1.326$. This uniform state differs from the real-space self-consistent clean state used in the impurity and vortex comparisons. All other spectra and responses use real-space self-consistent sample fields. Spectra are evaluated with supercell Bloch integration, and the Josephson response additionally uses imaginary-frequency integration. Normal-conductance calculations are specified in Appendix~\ref{app:frequency}.

As in the main text, coordinate pairs denote $(x/a,y/a)$; Fe sites have integer coordinates, whereas the marked core plaquette centers have half-integer coordinates. The $64\times32$ Fe cell has two orbitals per Fe and a reduced
$4N$-dimensional singlet BdG matrix. The main model parameters are
$(t_1,t_2,t_3,t_4,t_5,t_6)=(-1,0.08,1.35,-0.12,0.09,0.25)$,
$U=J_H=0$, $J_2=1.2$, target density $1.75$ per Fe, and
$T=10^{-4}$. No explicit orbital splitting, hopping strain,
Zeeman or spin--orbit term is applied.

The self-consistent clean and nonmagnetic-impurity $V=0.02,0.05,0.10$ states have
$\mu\simeq-1.324$; the two-vortex state has $\mu\simeq-1.325$.
The chemical potentials are displayed to three decimal places for presentation; spectral calculations use the full numerical values determined self-consistently. The self-consistent field updates are evaluated at zero supercell twist. Integration meshes of $12\times24$ and, for selected contacts, $24\times48$ Bloch
points integrate the spectra of the fixed self-consistent fields. They improve integration of observables
without establishing Bloch-mesh convergence of the self-consistent solution itself.
LDOS/QPI and Josephson maps use a $12\times24$ supercell Bloch mesh.
Selected contacts are compared with $24\times48$ points and 400/800-node
complete-frequency quadratures. The frequency-integration and normal-contact checks are given in Appendix~\ref{app:frequency}. The residual tolerance for self-consistency is $10^{-5}$ at zero field and $10^{-4}$ for vortex states. The additional vortex states in Figs.~\ref{fig:A2} and \ref{fig:A3} include solutions that reach the iteration cap with residuals below $10^{-4}$ but above the zero-field $10^{-5}$ tolerance.

For each of the three single-impurity potentials, the impurity is located at $(31,15)$. The off-center site
is $(29,13)$, whereas the impurity LDOS neighbor is $(29,15)$ and a separate
additional spectral neighbor is $(30,15)$. Clean contact checks include
$(31,15)$ and $(30,15)$; these are distinct from the off-center site $(29,13)$.
The clean-contact responses at these two sites agree to the reported precision. Vortex sampling
sites are C1$=(16,8)$ and N1$=(14,6)$; the plaquette cores are
$(15.5,7.5)$ and $(48.5,24.5)$. The two-impurity control has $V=0.02$ at
each of $(16,8)$ and $(48,24)$.

For each impurity potential, the NNN singlet pair-field components on all bonds were compared with their clean counterparts. None shows a relative $\pi$
shift. Maximum individual-bond magnitude changes are $0.4180\%$,
$1.004\%$ and $2.070\%$. They are distinct from the displayed
mean-amplitude contrast.

\section{Josephson response and tunneling strength}
\label{app:frequency}
\setlength{\abovedisplayskip}{5pt}
\setlength{\belowdisplayskip}{5pt}
\label{app:integration}
\label{app:contact}
\label{app:verification}
\par\noindent\begin{minipage}{\columnwidth}
\makeatletter\def\@captype{figure}\makeatother
\centering
\makebox[\linewidth][c]{\includegraphics[width=246bp]{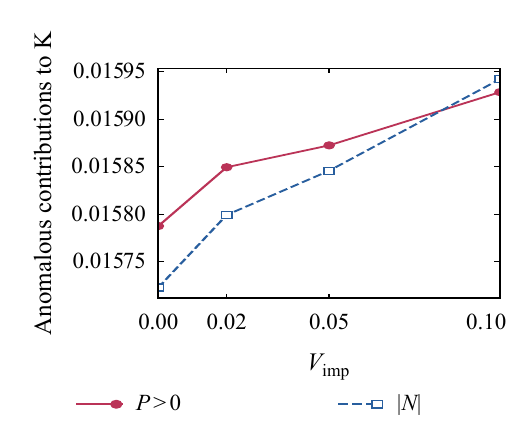}}
\caption{\label{fig:A14}Positive and negative anomalous spectral sums at the impurity center $(31,15)$. The positive sum $P$ and magnitude $|N|$ of the negative sum are each about $1.6\times10^{-2}$. The positive and negative anomalous spectral contributions nearly cancel, leaving a small net Josephson coupling. At the impurity center for $V_{\rm imp}=0.10$, both orbital contributions to the full-spectrum coupling are negative, so the compensation occurs between eigenstate contributions rather than opposite-sign final orbital sums. The sums use $W_0(E,\Delta_t)$ and the phase convention of Eq.~\eqref{eq:Fconvention}.}
\end{minipage}\par\addvspace{16pt}

All contact quantities use one fixed common phase reference and the same tip--sample phase reference. The sign of the real coupling is therefore defined relative to this common reference, rather than an independently chosen phase at each site. A consistent change of gauge transforms the anomalous coefficient and tunneling phase together and leaves the current unchanged.

At finite temperature, the Josephson response would use
\[
K_{i\alpha}(T)=T\sum_m
\frac{\Delta_t F_{i\alpha}(i\omega_m)}
{\sqrt{\omega_m^2+\Delta_t^2}},
\qquad \omega_m=(2m+1)\pi T.
\]
The reported Josephson response uses the continuous $T\to0$ limit, with sample fields held at their self-consistent values obtained at $T=10^{-4}$. Using the anomalous spectral weights
$R_{i\alpha n}^{\bm\kappa}=-u_{i\alpha n}^{\bm\kappa}
v_{i\alpha n}^{\bm\kappa *}$ in the phase convention of Eq.~\eqref{eq:Fconvention},
\[
K_{i\alpha}=\frac1{N_k}\sum_{\bm\kappa n}
 R_{i\alpha n}^{\bm\kappa}
 W_0(E_{n\bm\kappa},\Delta_t),
\]
with
\[
W_0(E,\Delta_t)=-\frac E\pi\int_0^\infty
\frac{\Delta_t\,d\omega}
{(E^2+\omega^2)\sqrt{\omega^2+\Delta_t^2}}.
\]
For $x=|E|/\Delta_t>0$,
\[
W_0(E,\Delta_t)=-\frac{\sgn E}{\pi}
\begin{cases}
\dfrac{\arccos x}{\sqrt{1-x^2}},&0<x<1,\\[2mm]
1,&x=1,\\[1mm]
\dfrac{\operatorname{arcosh}x}{\sqrt{x^2-1}},&x>1.
\end{cases}
\]
At exactly $E=0$ the original integrand defines $W_0(0,\Delta_t)=0$;
the one-sided $E\to0$ limits need not coincide. All maps and cumulative curves use this complete-frequency weight.

A cumulative curve includes states with $|E_{n\bm\kappa}|\le E_c$ after
evaluating their complete-frequency weights. Including the entire sample spectrum recovers the total Josephson coupling $K$. Direct nonmagnetic-impurity sums on a common
cutoff grid reproduce the independently evaluated full-spectrum Josephson coefficients to a maximum absolute
difference $3.65\times10^{-17}$.

The Josephson coefficient is obtained by integration over the full positive imaginary-frequency axis, using the self-consistent pair potential and densities. For $0\le\omega\le32$, we use 32-point Gauss--Legendre quadrature in each of the 11 intervals defined by the following imaginary frequencies: $0$, $0.02$, $0.05$, $0.1$, $0.2$, $0.5$, $1$, $2$, $4$, $8$, $16$ and $32$. The contribution from $\omega>32$ is evaluated with 48 quadrature points using $\omega=32/v$, $0<v<1$, with Jacobian $32/v^2$. This gives 400 quadrature points in total; doubling both point counts gives the 800-point frequency-integration check. The integration includes the tail beyond 32 rather than imposing a hard cutoff. The analytic weight $W_0$ above provides an independent check of the nonmagnetic-impurity couplings.

The zero-field eigenvectors are expressed in the contact phase convention using
$\mathcal U=\operatorname{diag}(1,1,-1,-1)$, giving
$K=6.471\times10^{-5}>0$ at $\Delta_t=0.20$. The sign convention is
unchanged across sites and tip gaps.

In the order $(e_{xz},e_{yz},h_{xz},h_{yz})$, the equivalent conversion
$D_i=\operatorname{diag}(i,i,-i,-i)$ acts on the full Green matrix as
$\bar G_i\mapsto D_i\bar G_iD_i^\dagger$. Vortex calculations retain the fixed
magnetic site factors. The minus sign in $F_\alpha=-[\bar G_s]_{e_\alpha h_\alpha}$
also enters the spectral residue and weak-tunneling formula; omitting it reverses the sign of the sine coefficient for the stated tip phase.

The full $12\times24$ nonmagnetic-impurity spatial Josephson coefficients and finer-contact values differ
by less than $8.4\times10^{-18}$; the two selected vortex contacts differ by at
most $1.87\times10^{-10}$. At the ten tip gaps, the largest 400/800-node
difference is $4.39\times10^{-18}$. These are Bloch-mesh and frequency-integration checks on fixed fields; cell-size and finer-mesh
convergence of the self-consistent solution are not established.

For the Josephson current beyond the weak-tunneling limit, both electrodes are described by fixed pair potentials. The tip normal DOS is
$\nu_t=1/\pi$ per spin in units of $1/t$. For real amplitudes
$\tau_\alpha=\sqrt{\Gamma_cw_\alpha}$,
$\Gamma_\alpha=\pi\nu_t|\tau_\alpha|^2=w_\alpha\Gamma_c$. The plotted scale is
the total $\Gamma_c=\Gamma_{xz}+\Gamma_{yz}$; each channel has half that value.
It is not a transmission eigenvalue or a spectral broadening.

For each tip orbital,
\[
g_t^{(\alpha)}(i\omega,\varphi)=
-\frac1{\sqrt{\omega^2+\Delta_t^2}}
\begin{pmatrix}i\omega&\Delta_te^{-i\varphi}\\
\Delta_te^{i\varphi}&i\omega\end{pmatrix}.
\]
The tip condensate phase is $-\varphi$. Assemble both channels in the order
$(e_{xz},e_{yz},h_{xz},h_{yz})$ and define
\begin{align}
W&=\sqrt{\Gamma_c}\operatorname{diag}
(\sqrt{w_{xz}},\sqrt{w_{yz}},-\sqrt{w_{xz}},-\sqrt{w_{yz}}),\nonumber\\
\Sigma_t&=Wg_tW^\dagger,\qquad M_i=1-\bar G_{s,i}\Sigma_t.
\end{align}

The current at finite tunneling strength uses the full local $4\times4$ orbital--Nambu Green matrix $\bar G_{s,i}$. The Bloch average is taken before evaluating $M_i^{-1}$, with the sample fields fixed and $\Delta_t=0.20$.
The total couplings are $\Gamma_c=10^{-6}$, $3\times10^{-6}$, $10^{-5}$, $3\times10^{-5}$, $10^{-4}$, $3\times10^{-4}$, $10^{-3}$, $0.003$, $0.01$, $0.03$ and $0.1$.
The weak-tunneling coefficients are evaluated at tip gaps
$0.16,0.18,0.185,0.19,0.195,0.20,0.205,0.21,0.22,0.24$ in a single unchanged
phase convention. The CPRs beyond the weak-tunneling limit are not recomputed at each of these gaps.
The equivalent amplitude--phase representation is
\[
I_i(\varphi)=\sum_{m\ge1}I_i^{(m)}
\sin[m\varphi+\delta_i^{(m)}].
\]

The current harmonics are obtained from $a_n=2\langle I\sin(n\varphi)\rangle$ and $b_n=2\langle I\cos(n\varphi)\rangle$, using $\varphi_j=2\pi j/128$ for $j=0,\ldots,127$ and retaining $n=1,\ldots,8$. The coefficient $a_1$ is signed, while $H_n=(a_n^2+b_n^2)^{1/2}$ is nonnegative. Dimensionless harmonic coefficients are converted to current units by multiplying by $2et/\hbar$. All normalized curves are divided by the positive weak-tunneling amplitude for the same sample state, contact site and tunneling strength:
\[
 I_{c,w}(i,\Gamma_c)=\frac{2et}{\hbar}\,2\Gamma_c|K_i|,
\]
rather than the maximum of the full CPR. The denominator is positive for
the displayed contacts and vanishes at $K_i=0$, where the ratios are
undefined. The deviation of the CPR from the weak-tunneling limit is
\[
 \epsilon_{\rm CPR}=\frac{\max_\varphi|I_{\rm full}(\varphi)-I_w(\varphi)|}
 {I_{c,w}}.
\]
The maximum Josephson current magnitude and $\epsilon_{\rm CPR}$ are evaluated at the same phase differences.
The sign of $a_1$ alone does not determine the minimum of the complete junction free energy. The self-consistent fields remain unchanged in the coupling scan at $\Delta_t=0.20$.
Finite-bias multiple Andreev reflection and tip-induced changes of the
fields are outside this equilibrium calculation.

The CPR deviation includes changes in the $m=1$ harmonic and in higher harmonics. The higher-harmonic ratio contains harmonics $n=2,\ldots,8$, normalized by $H_1$.

For $V_{\rm imp}=0.02$, the weak-tunneling couplings are positive at both $(31,15)$ and $(29,13)$, while the CPR changes with tunneling strength (Fig.~\ref{fig:A13}).

For the clean contact and the impurity-center contact at $V_{\rm imp}=0.10$, we compare the positive and negative weak-tunneling sine coefficients, respectively, with the corresponding full CPRs at the same tunneling strength.

At $V_{\rm imp}=0.10$, the off-center impurity contact $(29,13)$ has a negative weak-tunneling Josephson coefficient. At the Fe site C1$=(16,8)$ in the vortex state, the Josephson current has a phase offset in the common tip--sample phase reference (Fig.~\ref{fig:A10}). The CPR at N1$=(14,6)$ samples the response farther from the core than C1 (Fig.~\ref{fig:A11}(a)); the maximum Josephson current magnitudes at all five contacts are compared with their weak-tunneling amplitudes in Fig.~\ref{fig:A11}(b).

The normal conductance is calculated from the full two-orbital normal Green matrix. Write $g=g_N(E+i\eta_N)$ for the local electron Green matrix used to calculate the normal conductance,
$A_N=-(g-g^\dagger)/(2i)$,
$B=\sqrt{\Gamma_c}\operatorname{diag}(\sqrt{w_{xz}},\sqrt{w_{yz}})$,
and $\Sigma_N=-iB^2$. With the positive Hermitian square root of $A_N$,
\begin{align}
\mathcal A&=2B(1-g\Sigma_N)^{-1}A_N^{1/2},\nonumber\\
G_N/G_0&=\operatorname{Tr}(\mathcal A\mathcal A^\dagger),\qquad G_0=2e^2/h.
\end{align}
The plotted zero-bias conductance evaluates $E=0$ at the specified $\eta_N$.
Its weak limit is $4\pi\Gamma_c\nu_NG_0$ with
$\nu_N=\operatorname{Tr}(wA_N)/\pi$ per spin. With sample pairing set to zero, the normal Hamiltonian and chemical potential are retained from the corresponding superconducting state. The resulting density is not constrained to the target filling.
For normal conductance, $\eta_N$ is an imposed spectral broadening and does not enter the imaginary-frequency integrals for $K$ or the equilibrium CPR. With the normal Hamiltonian, chemical potential and positions fixed on the $24\times48$ Bloch mesh, reducing $\eta_N$ from $0.0015$ to $0.00075$ changes the normal DOS by $0.570\%$, $0.651\%$ and $0.480\%$ at clean A, the $V_{\rm imp}=0.10$ impurity center and the off-center site.

For the same reduction from $\eta_N=0.0015$ to $0.00075$, the weak-limit current--resistance products decrease by $0.573\%$, $0.655\%$ and $0.483\%$; each percentage uses the corresponding smaller-broadening value as denominator. Of the tested broadenings $0.003,0.0015,0.00075$, we use the smallest without establishing the zero-broadening limit. Independently, mesh refinement at $\eta_N=0.00075$ changes the zero-field products by $3.46$--$4.31\%$. The vortex normal DOS changes by $7.082\%$ between the last two broadenings, relative to the $0.00075$ value; no absolute vortex $I_cR_N$ or its zero-broadening limit is determined. The normal conductance and weak-limit products are shown in Fig.~\ref{fig:A8}.

\begin{widetext}
\noindent\begin{minipage}[t]{0.485\textwidth}
\makeatletter\def\@captype{figure}\makeatother
\centering
\includegraphics[width=0.94\columnwidth]{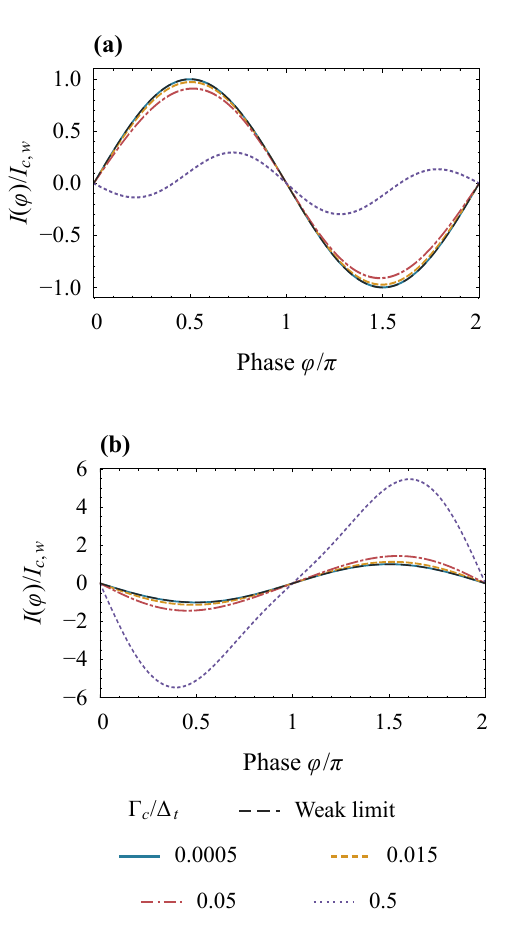}
\caption{\label{fig:A9}Equilibrium CPR at (a) clean A$=(31,15)$ and (b) the $V=0.10$ impurity center at $(31,15)$. All currents at finite tunneling strength are divided by the weak-tunneling amplitude at the same tunneling strength, rather than by their full-current maximum. The couplings denote total $\Gamma_c/\Delta_t$ with equal orbital weights and $\Delta_t=0.20$. At $0.015$ the CPR deviations are $2.987\%$ and $13.99\%$. The negative impurity-site coupling reverses the weak-tunneling current at a given phase difference.}
\end{minipage}\hfill
\begin{minipage}[t]{0.485\textwidth}
\makeatletter\def\@captype{figure}\makeatother
\centering
\includegraphics[width=0.94\columnwidth]{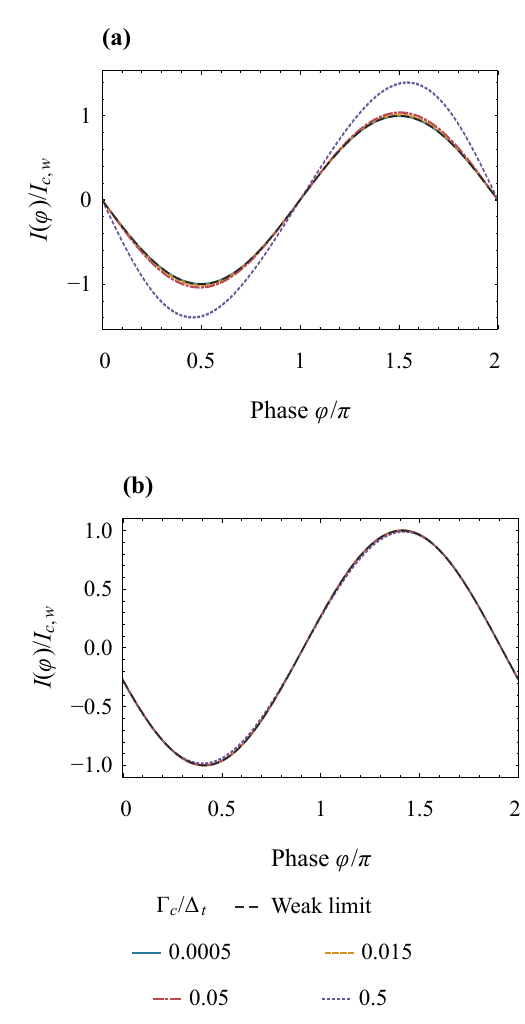}
\caption{\label{fig:A10}Position-dependent CPR at (a) the $V=0.10$ impurity off-center contact $(29,13)$ and (b) site C1$=(16,8)$ in the vortex state. Normalization and couplings match Fig.~\ref{fig:A9}. The vortex CPR has a phase offset in the common tip--sample phase reference. At $\Gamma_c/\Delta_t=0.5$, the CPR deviations are $42.14\%$ and $2.50\%$.}
\end{minipage}
\end{widetext}

\begin{widetext}
\noindent\begin{minipage}[t]{0.485\textwidth}
\makeatletter\def\@captype{figure}\makeatother
\centering
\includegraphics[width=0.72\columnwidth]{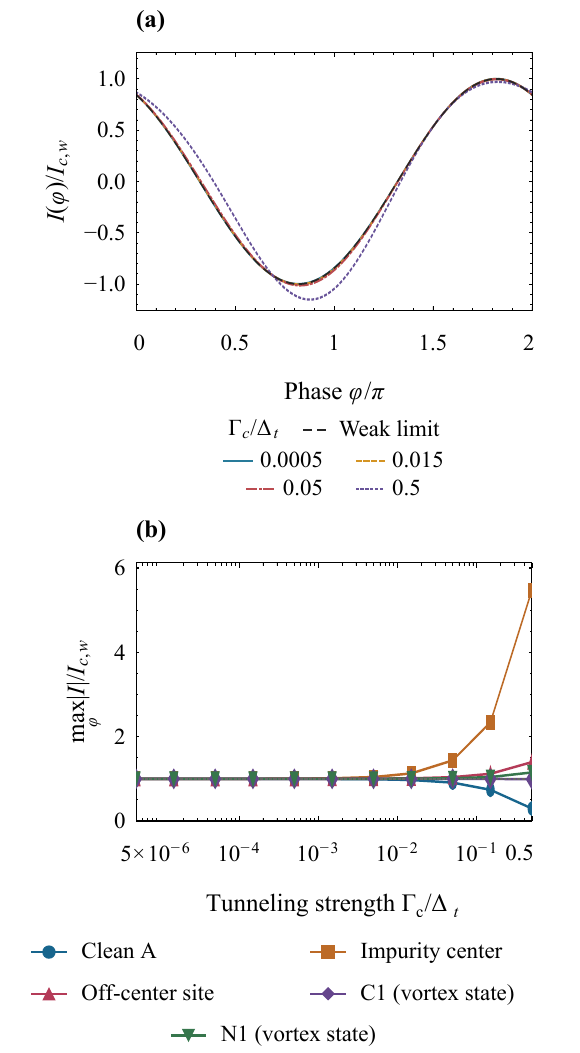}
\caption{\label{fig:A11}(a) CPR at site N1$=(14,6)$ in the vortex state, with the tip--sample phase reference and current normalization of Fig.~\ref{fig:A9}. At $\Gamma_c/\Delta_t=0.5$, the CPR deviation is $22.05\%$. (b) The maximum Josephson current magnitude, normalized by the weak-tunneling amplitude at the same tunneling strength, for the five contacts specified in the preceding text. The dashed line at unity shows the weak-tunneling prediction. The tunneling-strength axis in (b) is logarithmic.}
\end{minipage}\hfill
\begin{minipage}[t]{0.485\textwidth}
\makeatletter\def\@captype{figure}\makeatother
\centering
\includegraphics[width=0.72\columnwidth]{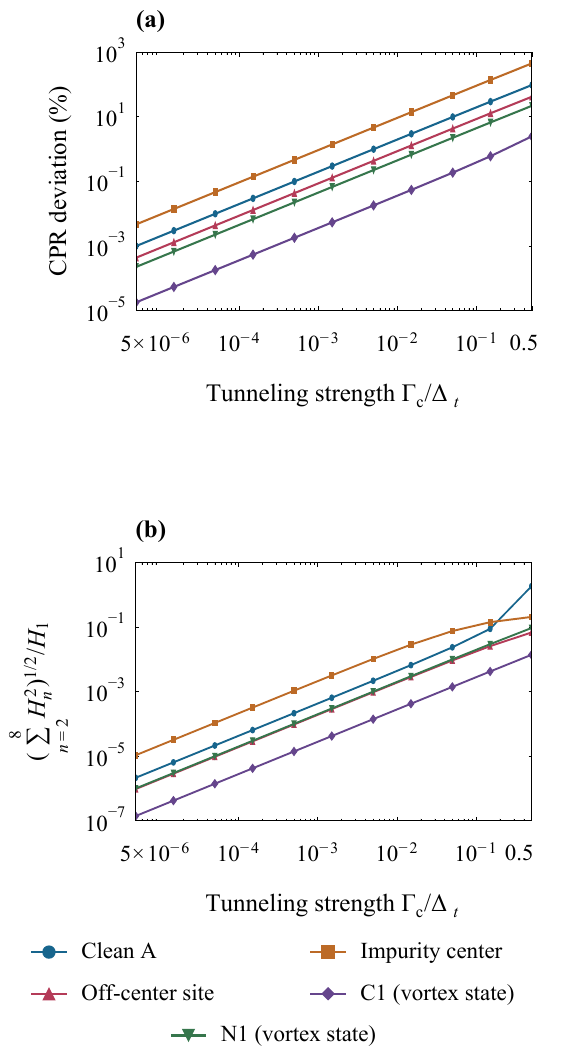}
\caption{\label{fig:A12}Deviations from the weak-tunneling current and higher Josephson harmonics for the five contacts defined in Fig.~\ref{fig:A11}. (a) The maximum deviation from the weak-tunneling current, normalized by the weak-tunneling amplitude and expressed as $100\epsilon_{\rm CPR}$. (b) The combined amplitude of higher Josephson harmonics, $n=2,\ldots,8$, relative to the fundamental amplitude $H_1$, retaining sine and cosine components. Changes in the fundamental harmonic also contribute to (a). Both panels use logarithmic axes.}
\end{minipage}
\par\vspace{10pt}
\noindent\begin{minipage}{\textwidth}
\makeatletter\def\@captype{figure}\makeatother
\centering
\includegraphics[width=400bp]{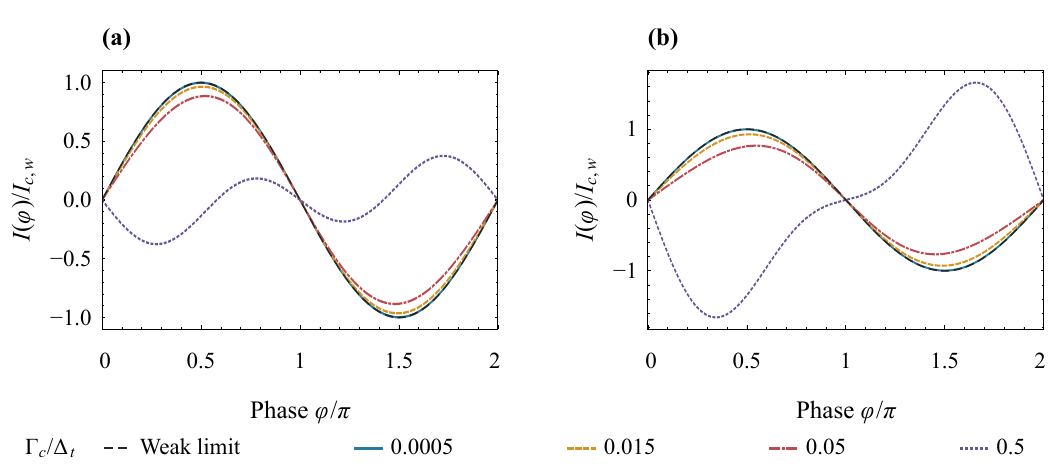}
\caption{\label{fig:A13}CPRs for $V=0.02$ at (a) the impurity center $(31,15)$ and (b) the off-center site $(29,13)$. The weak-tunneling Josephson couplings are positive at both sites. Currents are normalized by the weak-tunneling amplitude at the same tunneling strength, with the phase reference used in Fig.~\ref{fig:A9}. At $\Gamma_c/\Delta_t=0.015$, CPR deviations are $3.840\%$ and $7.929\%$.}
\end{minipage}
\par\vspace{10pt}
\noindent\begin{minipage}{\textwidth}
\makeatletter\def\@captype{figure}\makeatother
\centering
\includegraphics[width=450bp]{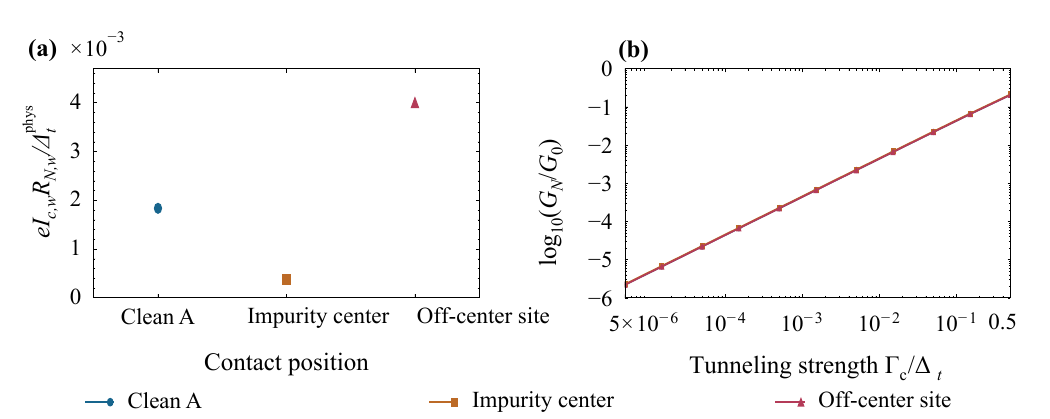}
\caption{\label{fig:A8}Normal conductance and the weak-tunneling current--resistance product. Clean A and impurity C are at $(31,15)$; impurity N is at $(29,13)$. (a) Weak-limit $eI_{c,w}R_{N,w}/\Delta_t^{\rm phys}=0.001838,0.0003753,0.004007$ for clean A and the $V_{\rm imp}=0.10$ impurity contacts C and N. Both factors in each product are weak-limit quantities. (b) Base-ten logarithm of the normal conductance, $\log_{10}(G_N/G_0)$, versus total $\Gamma_c/\Delta_t$, with $G_0=2e^2/h$. The horizontal axis is logarithmic. The normal conductance uses $\Delta=0$ and the normal Hamiltonian and chemical potential of the corresponding superconducting state, with $\eta_N=0.00075$. Refining the Bloch integration changes these products by $4.31\%$, $3.46\%$, and $3.59\%$, respectively.}
\end{minipage}
\end{widetext}

\section{Finite onsite repulsion}
\label{app:finiteu}
We compare $U=0$ and $1$ on the nonmagnetic, spin-symmetric branch at $J_H=0$, $J_2=1.2$, $n=1.75$ and $V_{\rm imp}=0.05$. The interaction convention is $U'=U-2J_H$. The diagonal Hartree term in the electron normal block is
\[
h^{\rm H}_{i\alpha\sigma}=U n_{i\alpha\bar\sigma}
 +(U-2J_H)n_{i\bar\alpha\bar\sigma}
 +(U-3J_H)n_{i\bar\alpha\sigma},
\]
where $\bar\alpha$ denotes the other orbital. With $J_H=0$, $U'=U$. We impose $n_{i\alpha\uparrow}=n_{i\alpha\downarrow}$ while allowing spatial and orbital charge redistribution. The clean and impurity states are separately self-consistent, with $\mu$ adjusted to maintain the filling; its definition includes the Hartree shift. For $U=1$, the clean and impurity chemical potentials are both approximately $-0.011$. The respective maximum field residuals are $6.76\times10^{-6}$ and $5.27\times10^{-6}$, and filling deviations are below $7.53\times10^{-7}$.

The clean density is spatially uniform and orbital symmetric, so $h^{\rm H}=3Un/4$ is absorbed by the chemical potential. The independently obtained self-consistent clean states at $U=0$ and $U=1$ differ in $\mu-h^{\rm H}$ by $2.25\times10^{-6}$ and in their NNN bond amplitudes by less than $7.90\times10^{-6}$. The clean Josephson coefficients differ by $0.462\%$; this difference is not attributed to uniform Hartree charge feedback. The integration tests below hold the sample fields fixed and do not assess the difference between the two self-consistent clean states.

Table~\ref{tab:finiteu} uses $\Delta_t=0.20$, equal orbital weights, the full positive imaginary-frequency integral and the same sample phase convention. $K$ is the dimensionless Josephson coefficient; its conversion to current is unchanged.
\begin{table}[!htb]
\caption{\label{tab:finiteu}Self-consistent charge-feedback comparison. Ratios use the clean coefficient at the same $U$ and site.}
\begin{ruledtabular}
\begin{tabular}{clrr}
$U$ & State or position & $10^5\operatorname{Re}K$ & $K/K_{\rm clean}$\\
\hline
0 & Clean & 6.471 & 1\\
0 & $(31,15)$ & 2.672 & 0.4129\\
0 & $(29,13)$ & $-3.790$ & $-0.5857$\\
1 & Clean & 6.441 & 1\\
1 & $(31,15)$ & 4.017 & 0.6236\\
1 & $(29,13)$ & $-0.4305$ & $-0.06683$\\
\end{tabular}
\end{ruledtabular}
\end{table}
For $U=1$, changing the Bloch integration mesh from $12\times24$ to $24\times48$ changes $\operatorname{Re}K$ by at most $2.12\times10^{-18}$ for the clean coupling and $3.92\times10^{-18}$ for the two impurity contacts. On the finer mesh, the maximum 400/800-point frequency-integration difference is $9.49\times10^{-20}$; the 800-point and analytic frequency weights differ by at most $3.66\times10^{-19}$. These differences are far below the magnitude of the negative coefficient closest to zero in this comparison, $K=-4.30\times10^{-6}$, whose sign is unchanged. The self-consistent states are obtained at zero supercell twist and remain fixed during spectral integration; these comparisons test Bloch and frequency integration for those states, not Bloch-mesh convergence of the self-consistent solution.

For these nonmagnetic solutions, $U=1$ weakens the suppression of the impurity-site Josephson coupling and reduces the magnitude of the negative coupling at $(29,13)$ (Fig.~\ref{fig:finiteuSCF} and Table~\ref{tab:finiteu}). These results describe the two values of $U$, the nonmagnetic branch and the specified positions only.
\begin{widetext}
\noindent\begin{minipage}{\textwidth}
\makeatletter\def\@captype{figure}\makeatother
\centering
\includegraphics[width=440bp]{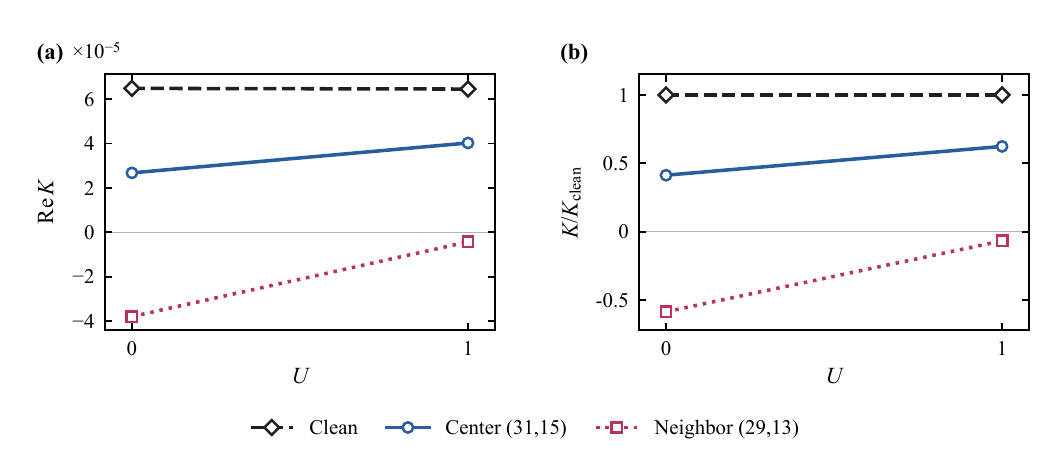}
\caption{\label{fig:finiteuSCF}Self-consistent charge-feedback comparison at $J_H=0$, $J_2=1.2$, $n=1.75$ and $V_{\rm imp}=0.05$. (a) Unnormalized $\operatorname{Re}K$ for the clean state and impurity contacts $(31,15)$ and $(29,13)$. (b) Ratios to the clean coefficient at the same $U$ and site. The tip gap is $\Delta_t=0.20$ with equal orbital weights and the same sample phase convention. Markers show the calculated values at $U=0$ and $U=1$; lines connect the two values. The $U=1$ impurity coefficients are evaluated for the self-consistent states using a $24\times48$ Bloch integration mesh.}
\end{minipage}
\end{widetext}

\section{Additional spectra and QPI}
\label{app:qpi}

The QPI amplitudes obey
$Q(\mathbf q,E)=N^{-1}\sum_i\delta\rho_{W,i}(E)
 e^{-i\mathbf q\cdot\mathbf r_i}$, with $N=N_xN_y$.
All QPI maps retain the same absolute Fourier normalization. At $E=0.194$, the clean
LDOS at the impurity center is $0.7888$. The ratios
$\delta\rho_W/\rho_{W,\mathrm{clean}}$ there are
$+0.0322\%$, $-0.340\%$ and $-1.143\%$ for
$V=0.02,0.05,0.10$, respectively. These local ratios differ from the
cell-normalized Fourier amplitudes.
For the two-impurity and two-vortex maps in Fig.~\ref{fig:A7},
$|Q(0)|=1.32\times10^{-5}$ and $0.134$, while the nonzero-$q$
maxima are $1.45\times10^{-4}$ and $0.0813$.
The logarithmic color scale in Fig.~\ref{fig:A7} has a lower limit of $10^{-9}$.

The displayed quasiparticle maps use spin-summed orbital LDOS weighted
by $(1/2,1/2)$, broadening $\eta_s=0.003$, and energy spacing $0.002$.
The principal peaks of the clean LDOS occur at $E=-0.196$ and $+0.194$ on this energy grid.
The LDOS and QPI calculations use the full numerical chemical potentials determined for the corresponding self-consistent sample states. Values quoted in the text are rounded to three decimal places for presentation only; for $V=0.10$, $\mu\simeq-1.324$.

The maps show $|Q|$ with the same $1/N$ Fourier normalization, without smoothing, rotational symmetrization or removal of $q=0$. The original $64\times32$ Fourier grid has steps $1/32$ and $1/16$ in $q_x/\pi$ and $q_y/\pi$, respectively. The line cut follows $q_y=0$. The Brillouin-zone boundary sample at $-1\equiv+1$ is split between opposite periodic half-cells without adding an independent $+1$ point. The principal coherence peaks set the energy references. The additional QPI maps are evaluated at the negative-energy principal coherence peak and the inner positive-energy maximum of the clean LDOS; the latter is not assigned to a separate gap.

The single-impurity, two-impurity and two-vortex states have different spatial LDOS modulations; the corresponding spectra and the QPI of the latter two states are compared in Fig.~\ref{fig:A7}. The two identical impurities are separated by a half-cell translation.

\noindent\begin{minipage}{\columnwidth}
\makeatletter\def\@captype{figure}\makeatother

\centering
\includegraphics[width=\linewidth]{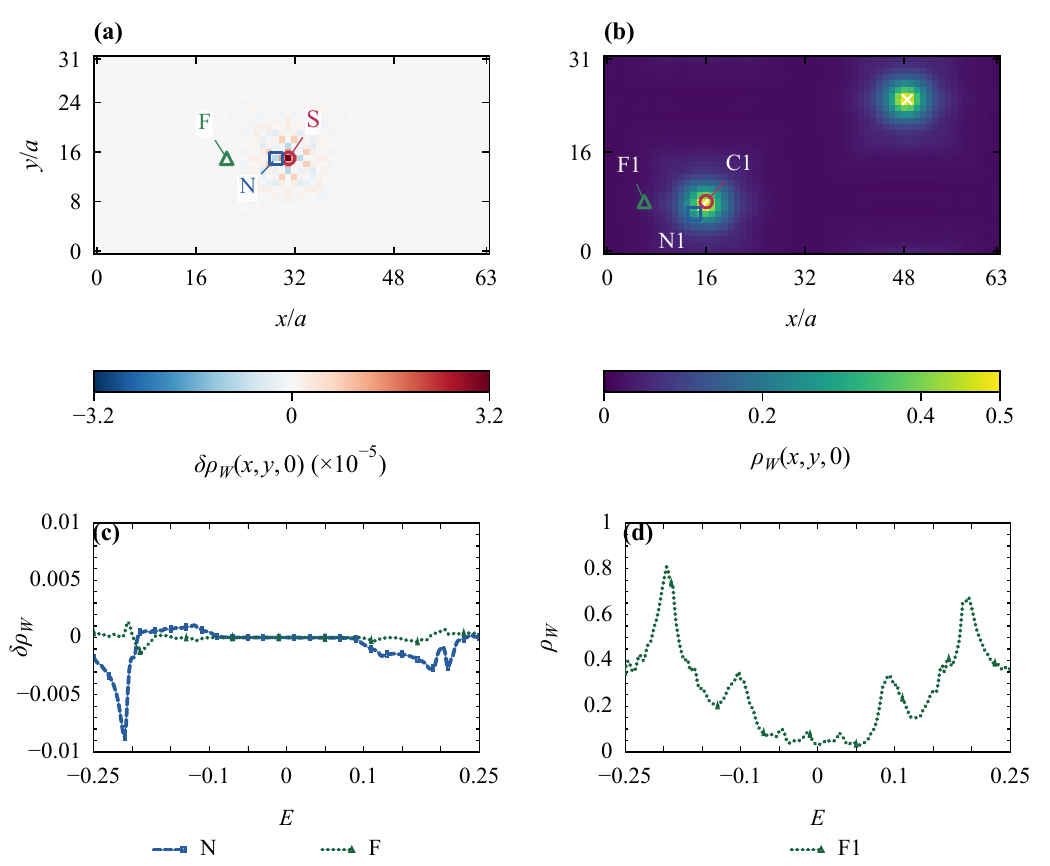}
\caption{\label{fig:A1}Local quasiparticle spectra near an impurity and a vortex. The weak nonmagnetic impurity has $V=0.02$. (a) Signed weak-impurity LDOS change at $E=0$, relative to the LDOS of the independently obtained clean superconducting state on the same Fe sublattice. (b) Total vortex LDOS at $E=0$; crosses mark plaquette cores. (c) Weak-impurity difference spectra at N$=(29,15)$ and F$=(21,15)$; the center spectrum is shown in Fig.~\ref{fig:scalarQPI}(f). (d) Total vortex spectrum at the remote site F1$=(6,8)$; the C1, N1 and clean spectra are shown in Fig.~\ref{fig:vortex}(c). The spectrum at F1 is from the same finite-field vortex state.}

\end{minipage}
\par\addvspace{12pt}

The independently obtained real-space clean superconducting state uses its self-consistently determined $\mu$ specified above. The superconducting LDOS differs from the electron DOS used to calculate the normal resistance, for which only the sample pair potential is set to zero.

\begin{widetext}
\noindent\begin{minipage}{\textwidth}
\makeatletter\def\@captype{figure}\makeatother

\centering
\includegraphics[width=504bp]{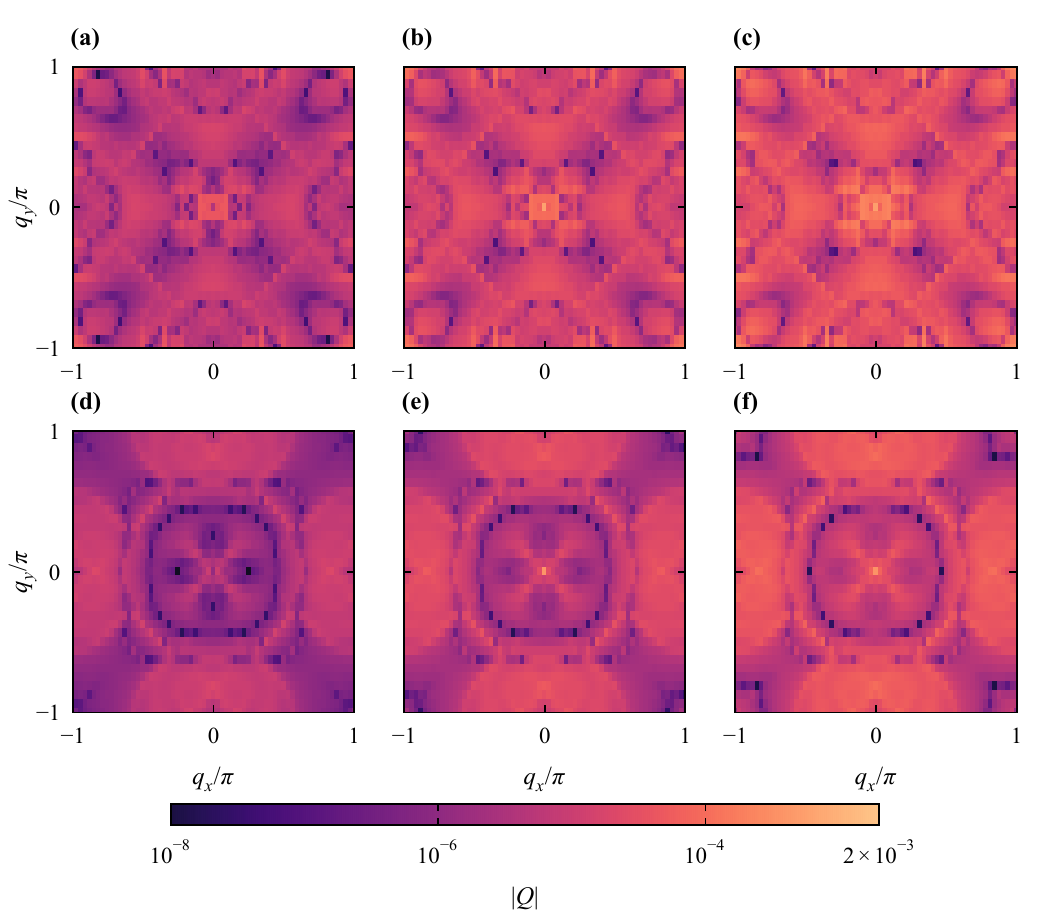}
\caption{\label{fig:A6}Energy dependence of nonmagnetic-impurity QPI. Panels (a--c) show $|Q|$ at $E=-0.196$, the negative principal clean coherence peak. Panels (d--f) show $|Q|$ at the inner positive-energy maximum of the clean LDOS, $E=+0.124$, which is not independently identified as a second gap. Columns correspond to $V=0.02,0.05,0.10$. The impurity potentials, clean superconducting state and absolute color scale match Fig.~\ref{fig:scalarQPI}. The corresponding $E=+0.194$ momentum cut is shown in Fig.~\ref{fig:scalarQPI}(e). The shared color scale is logarithmic.}
\end{minipage}
\end{widetext}

\begin{widetext}
\noindent\begin{minipage}{\textwidth}
\makeatletter\def\@captype{figure}\makeatother
\centering
\includegraphics[width=504bp]{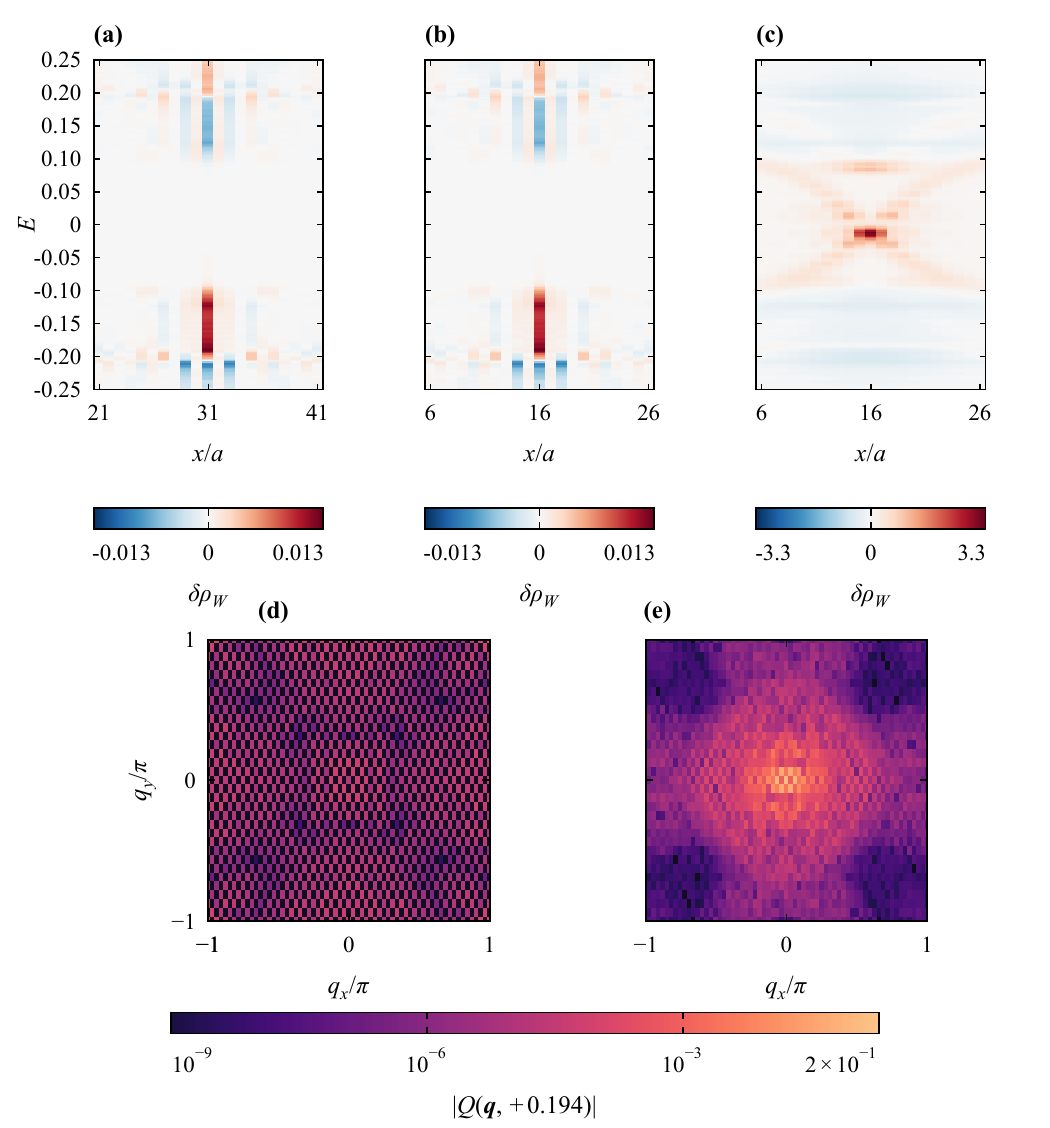}
\caption{\label{fig:A7}LDOS and QPI in impurity and vortex states. Each impurity center has $V=0.02$; the pair is at $(16,8)$ and $(48,24)$. Panels (a--c) show signed LDOS changes on $y/a=15,8,8$ for one nonmagnetic impurity, two nonmagnetic impurities and the two-vortex state, respectively, with separate indicated impurity and vortex ranges. Panels (d,e) show $|Q(\mathbf q,+0.194)|$ for two nonmagnetic impurities and the two-vortex state on a shared absolute scale, using the independently obtained clean superconducting state on the A/B Fe sublattices. The single-impurity QPI is shown in Fig.~\ref{fig:scalarQPI}(b). Suppression at selected wave vectors follows from destructive interference between the two equivalent impurities when the state preserves the half-cell translation symmetry. The color scales in (a--c) are linear, whereas (d,e) share a logarithmic color scale.}
\end{minipage}
\end{widetext}

\begin{widetext}
\noindent\begin{minipage}{\textwidth}
\makeatletter\def\@captype{figure}\makeatother
\appcols{
\section{Vortex pair fields and phase winding}
\label{app:phase}
The self-consistent vortex states are compared for opposite field directions, different gauges and different initial pair fields (Figs.~\ref{fig:A2} and \ref{fig:A3}), using the self-consistency conditions specified in Appendix~\ref{app:source}.
}\par\vspace{6pt}
\centering
\includegraphics[width=440bp]{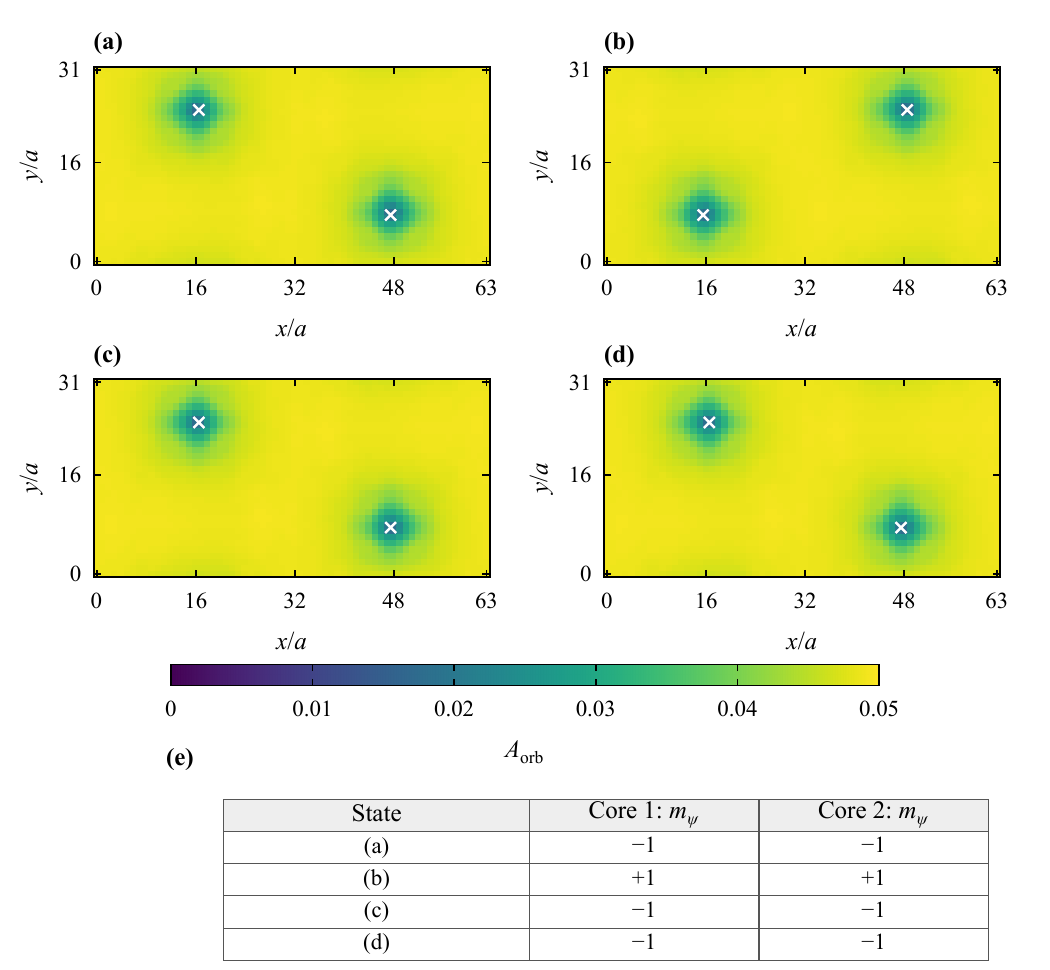}
\caption{\label{fig:A2}Vortex pair-field amplitudes for opposite field directions and different gauges. (a) Symmetric gauge, positive field, starting from an initial pair field with imposed vortex phase winding and core-amplitude suppression. (b) Reversed field starting from an initial pair field with no imposed vortex winding. (c,d) Positive-field states in the Landau gauges $A_y=Bx$ and $A_x=-By$, respectively, starting from an initial pair field with no imposed vortex winding. Maps show the orbital-averaged bond-pair amplitude $A_{\rm orb}$, which differs from the mean of the separate orbital bond magnitudes $A$ in Fig.~\ref{fig:vortex}(a). (e) Each core has the indicated phase-winding number on each square contour of side $2a,4a,6a,8a,10a,12a$. Each state is obtained from a separate self-consistent calculation, with the tolerance specified in Appendix~\ref{app:source}.}
\end{minipage}
\end{widetext}

A local field $X_i$, such as $\psi_i$ or the Josephson coefficient, transforming with charge $2e$ has
a link-covariant phase step
\[
d_{ij}[X]=\Arg[X_jX_i^*U_{ij}^2],
\qquad \Arg z\in(-\pi,\pi].
\]
For a closed loop $\mathcal C$ that does not wrap around the periodic cell, we use
\begin{equation}
m_X(\mathcal C)=\frac1{2\pi}
\left[\sum_{(ij)\in\mathcal C}d_{ij}[X]
-2\sum_{(ij)\in\mathcal C}\varphi_{ij}\right].
\label{eq:winding}
\end{equation}
The four-corner plaquette phase-winding index of the Josephson coefficient measures lattice-resolved circulation;
locating a continuous zero requires finer spatial sampling. Every periodic-boundary crossing includes the magnetic transition
phase in $U_{ij}$ and the corresponding charge-2e transformation of $X$.
Under $c_i\mapsto e^{i\chi_i}c_i$, one has
$U_{ij}\mapsto e^{i(\chi_i-\chi_j)}U_{ij}$ and
$X_i\mapsto e^{2i\chi_i}X_i$, leaving the edge product invariant.
Counterclockwise Fe-plane loops give index $-1$ for positive-field
condensate cores.

\begin{widetext}
\noindent\begin{minipage}{\textwidth}
\makeatletter\def\@captype{figure}\makeatother
\appcols{
The pair-field amplitude maps in Figs.~\ref{fig:A2} and \ref{fig:A3} use
\[
 A_{{\rm orb},i}=\frac14\sum_{\delta\in{\rm NNN}}
 \left|\frac{\Delta_{i,i+\delta}^{xz,\uparrow\downarrow}
 +\Delta_{i,i+\delta}^{yz,\uparrow\downarrow}}2\right|.
\]
It combines orbitals before taking the magnitude and differs from $A_i$.
}\par\vspace{3pt}
\centering
\includegraphics[width=450bp]{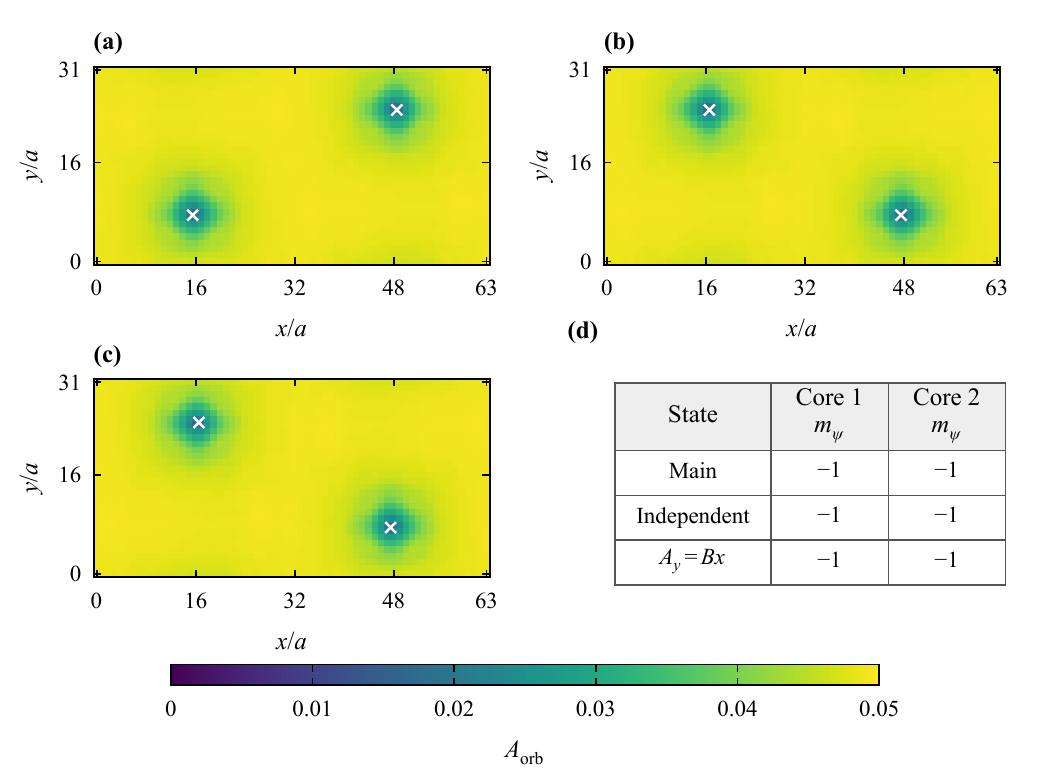}
\caption{\label{fig:A3}Self-consistent vortex states for different gauges and initial pair fields. (a) Main two-vortex state. (b) A second self-consistent state in the symmetric gauge. (c) A self-consistent state in the Landau gauge $A_y=Bx$. The initial pair fields in (a,b) have no imposed vortex winding. Panel (c) starts from a pair field with imposed vortex winding and core-amplitude suppression, and therefore differs in both the initial pair field and the gauge. All amplitude maps use the same $A_{\rm orb}$ color scale and Fe-site coordinates. (d) Both cores have phase-winding number $-1$ on each contour specified in Fig.~\ref{fig:A2}.}
\end{minipage}
\end{widetext}

\begin{widetext}
\noindent\begin{minipage}{\textwidth}
\makeatletter\def\@captype{table}\makeatother
\centering
\caption{\label{tab:loopwinding}Common counterclockwise loops around vortex cores 1 and 2 at $(15.5,7.5)$ and $(48.5,24.5)$, respectively. $C_X$ is the link-covariant phase circulation, $F_{\mathcal C}$ the signed magnetic term, and $m_X=C_X-F_{\mathcal C}$ the flux-subtracted winding. $m_\theta$ is evaluated from the relative phase $\theta_i=\arg(K_i\psi_i^*)$. Entries show six decimal places; unrounded values satisfy the integer and consistency checks within $10^{-12}$.}
\begin{ruledtabular}
\begin{tabular}{ccrrrrrrr}
Core & $L/a$ & $A/a^2$ & $F_{\mathcal C}$ & $C_\psi$ & $C_K$ & $m_\psi$ & $m_K$ & $m_\theta$\\
\hline
1 & 2 & 4 & 0.003906 & -0.996094 & -0.996094 & -1.000000 & -1.000000 & 0.000000 \\
1 & 4 & 16 & 0.015625 & -0.984375 & 0.015625 & -1.000000 & 0.000000 & 1.000000 \\
1 & 6 & 36 & 0.035156 & -0.964844 & 3.035156 & -1.000000 & 3.000000 & 4.000000 \\
2 & 2 & 4 & 0.003906 & -0.996094 & -0.996094 & -1.000000 & -1.000000 & 0.000000 \\
2 & 4 & 16 & 0.015625 & -0.984375 & 1.015625 & -1.000000 & 1.000000 & 2.000000 \\
2 & 6 & 36 & 0.035156 & -0.964844 & 3.035156 & -1.000000 & 3.000000 & 4.000000 \\
\end{tabular}
\end{ruledtabular}
\end{minipage}
\end{widetext}

Table~\ref{tab:loopwinding} compares the condensate and Josephson-response windings on common loops around each vortex. On the smallest loops both fields have winding $-1$, whereas larger loops retain the same condensate winding but enclose additional winding of the Josephson response. The windings are evaluated from the full-frequency Josephson response shown in Figs.~\ref{fig:vortexJosephson} and \ref{fig:A4} and the local pair field $\psi_i$, using the same gauge and magnetic links. Define $C_X=(2\pi)^{-1}\sum_{\mathcal C}d_{ij}[X]$ and $F_{\mathcal C}=(2\pi)^{-1}2\sum_{\mathcal C}\varphi_{ij}$, so that $m_X=C_X-F_{\mathcal C}$. $C_X$ contains the enclosed-flux contribution, which is removed in the winding $m_X$.

\begin{widetext}
\noindent\begin{minipage}{\textwidth}
\makeatletter\def\@captype{figure}\makeatother
\centering
\includegraphics[width=450bp]{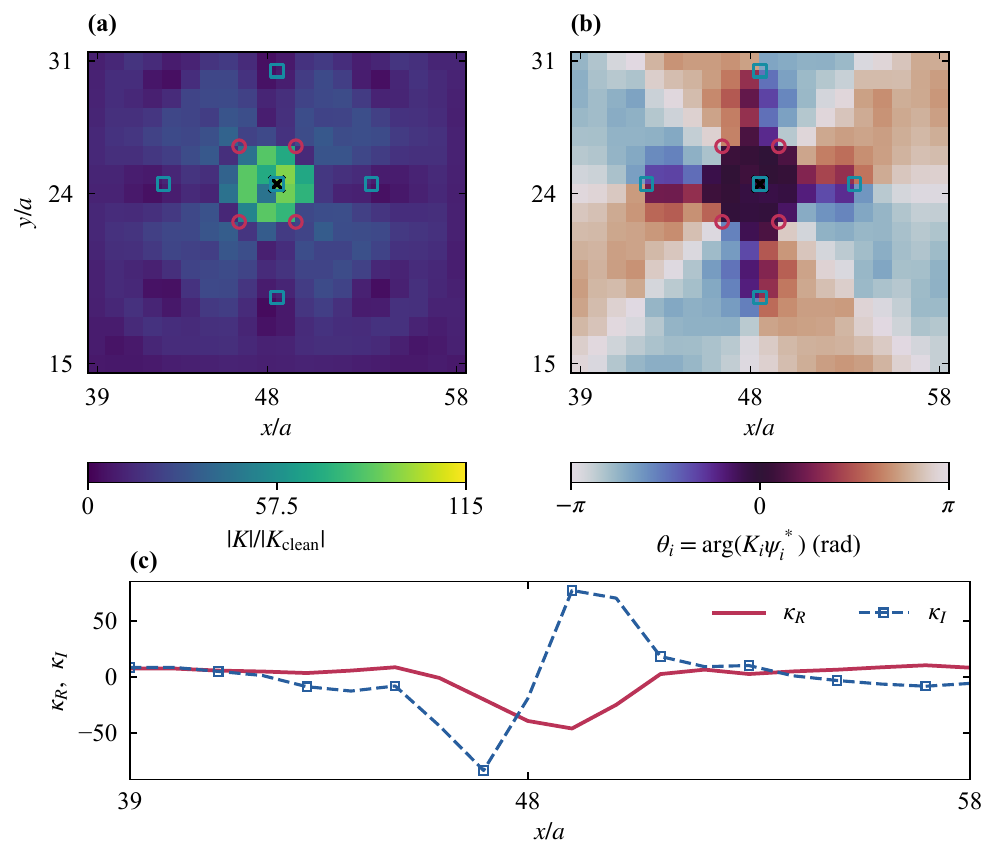}
\caption{\label{fig:A4}Local Josephson response near the second vortex core $(48.5,24.5)$. Panel (b) shows the phase relative to the condensate, $\theta_i=\arg(K_i\psi_i^*)$; the magnitude normalization, gauge and common phase reference are otherwise the same as in Fig.~\ref{fig:vortexJosephson}. The profile uses the Fe row $y/a=24$. The denominator $|K_{\rm clean}|$ is the coefficient magnitude of the impurity-free superconducting state at zero field. Circles and squares indicate selected plaquettes with Josephson-response windings $m_K=+1$ and $-1$, respectively, satisfying the frequency-integration criterion in Appendix~\ref{app:phase}.}

\end{minipage}
\end{widetext}

For the following dimensionless path coordinates, let $\ell=L/a$ be the side length in Fe-lattice units. The paths start at the integer Fe sites $(x_0,y_0)=(c_x-\ell/2,c_y-\ell/2)$, with $(c_x,c_y)=(15,7)$ or $(48,24)$ and even $\ell$. Their ordered corners are $(x_0,y_0)\to(x_0+\ell,y_0)\to(x_0+\ell,y_0+\ell)\to(x_0,y_0+\ell)\to(x_0,y_0)$, including every unit lattice step and the closing edge. This order is counterclockwise in the Fe plane and has signed area $+\ell^2a^2=+L^2$. These paths enclose, but are not exactly centered on, the half-integer vortex cores. Each edge uses the principal value in $(-\pi,\pi]$; the sequence of edge differences is not unwrapped.

The relative-phase winding $m_\theta=(2\pi)^{-1}\sum_{\mathcal C}\Arg[(K_j\psi_j^*)(K_i\psi_i^*)^*]$ is calculated from the relative phase $\theta_i=\arg(K_i\psi_i^*)$, without magnetic links. For the paths in Table~\ref{tab:loopwinding}, the relative-phase winding satisfies $m_\theta=m_K-m_\psi$; the sum of the edgewise branch differences $(d_{ij}[K]-d_{ij}[\psi]-d_{ij}[\theta])/(2\pi)$ vanishes within floating-point precision.

 We checked finite-frequency integrals using 96 and 192 frequency nodes with $W_c=8$ and $12$. For each contour $\mathcal C$, the edge margin is
$m_{\mathcal C}=\min_{(ij)\in\mathcal C}(\pi-|d_{ij}|)$.
The displayed plaquette phase windings require the largest edge-step change,
divided by this margin, to be below $0.25$. This condition tests the sensitivity of the plaquette winding to frequency integration. Using the complete-frequency Josephson coefficient leaves the branch choice for every displayed edge phase step unchanged.

At C1 and N1, contributions from different energy ranges partly cancel in the real and imaginary parts of the Josephson coupling (Fig.~\ref{fig:A16}). The cumulative couplings use the same tip weight as in Fig.~\ref{fig:spectralComp}. Including the full BdG spectrum gives the local coefficients used in the spatial Josephson maps.

\par\noindent\begin{minipage}{\columnwidth}
\makeatletter\def\@captype{figure}\makeatother
\centering
\includegraphics[width=0.94\columnwidth]{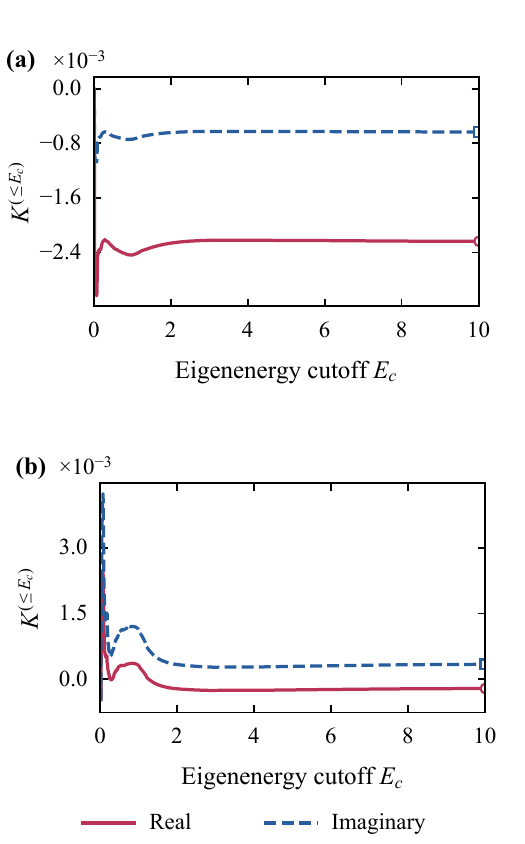}
\caption{\label{fig:A16}Energy-resolved contributions to the complex Josephson coupling at Fe sites (a) C1$=(16,8)$ and (b) N1$=(14,6)$ in the vortex state, in the fixed gauge and common phase reference. Real and imaginary components include states with $|E_n|\le E_c$ and the full energy-dependent Josephson weight for an $s$-wave tip with $\Delta_t=0.20$. Open symbols show the full-spectrum $K_i$ values at the same sites. The real and imaginary components of $K$ are shown without normalization by $K_{\rm clean}$. The corresponding spectral decomposition for the nonmagnetic-impurity states is shown in Fig.~\ref{fig:spectralComp}.}
\end{minipage}\par
On the tabulated paths the minimum sampled $|\psi|$ is $8.28\times10^{-3}$ and the minimum $|K|$ is $4.01\times10^{-4}$. The common-loop windings are unchanged in the frequency-integration checks, with the largest edge-phase change remaining below $0.029$ of the branch margin. Summing all elementary-plaquette windings inside each path reproduces $m_K$, $m_\psi$ and $m_\theta$. At the smallest paths $m_\theta=0$. Only plaquette windings satisfying this frequency-integration condition are marked. Their visible net winding can therefore differ from the total enclosed by a loop: for the core-1 $L=6a$ loop, the marked plaquettes sum to $+2$, whereas all enclosed plaquettes sum to $+3$, including the unmarked $+1$ plaquette at $(17.5,6.5)$. The windings refer to the sampled Fe-site fields; continuous zero positions are not determined.

The mean bond magnitude $A_i$, the local bond-pair-field magnitude $|\psi_i|$, and the local Josephson magnitude $|K_i|$ have different spatial profiles near core 1 (Fig.~\ref{fig:vortexprofiles}). Both pair-field magnitudes are suppressed near the core and increase outward, while the peaks and minima of $|K_i|$ depend on the cut direction. The magnitude $|\psi_i|$ also depends on the relative phases of the bond pair fields. At the ends of the plotted cuts, $A_i$ and $|\psi_i|$ remain below the clean value. No recovery length is extracted from these cuts.

To resolve the two diagonal bond directions at fixed orbital and sublattice, define $B_{i\delta}^{\alpha}=U_{i,i+\delta}\Delta_{i,i+\delta}^{\alpha,\uparrow\downarrow}$ in the same gauge as $\psi_i$, and
\[
\mathcal S_i^\alpha=\tfrac14(B_{i,\rm NE}^\alpha+B_{i,\rm SW}^\alpha+B_{i,\rm SE}^\alpha+B_{i,\rm NW}^\alpha),
\]
\[
\mathcal D_i^\alpha=\tfrac14(B_{i,\rm NE}^\alpha+B_{i,\rm SW}^\alpha-B_{i,\rm SE}^\alpha-B_{i,\rm NW}^\alpha).
\]

Then $\psi_i=(\mathcal S_i^{xz}+\mathcal S_i^{yz})/2$. The clean-state magnitudes are $|\mathcal S^\alpha|=0.05002t$ and $|\mathcal D^\alpha|=0.003718t$; the relative signs of the components depend on the orbital and Fe sublattice. Within $6.5a$ of either core, local maxima of $|\mathcal D^\alpha|$ exceed their same-orbital, same-sublattice clean values by approximately $25$--$31\%$. The quoted percentage changes compare $|\mathcal D_i^\alpha|$ with its clean value; $|\mathcal D_i^\alpha|$ is suppressed at other near-core sites. $\mathcal S_i^\alpha$ and $\mathcal D_i^\alpha$ describe the bond-direction components of the pair field. Their local changes do not by themselves establish an additional superconducting phase or explain the winding of the Josephson response.

\begin{widetext}
\noindent\begin{minipage}{\textwidth}
\makeatletter\def\@captype{figure}\makeatother
\centering
\includegraphics[width=450bp]{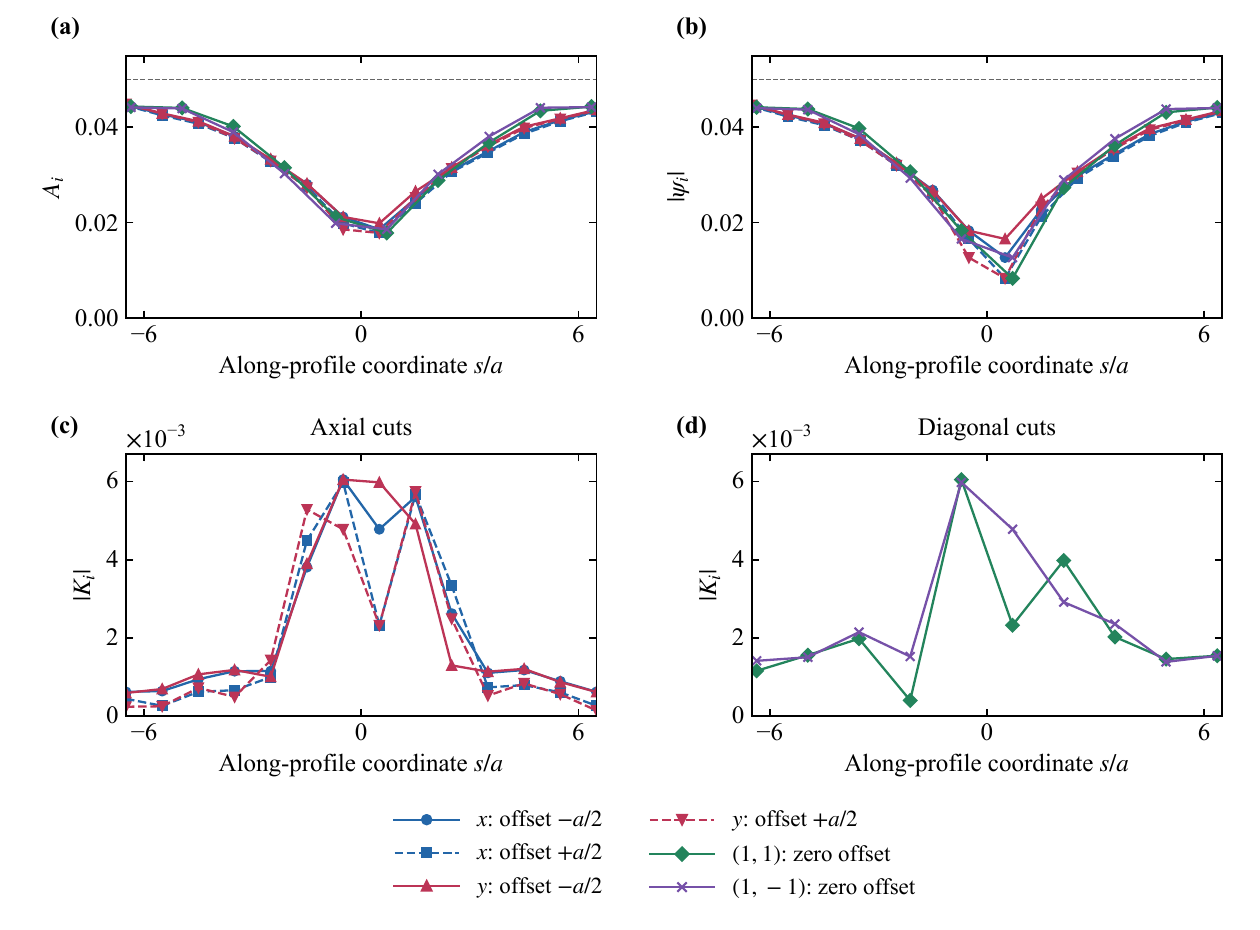}
\caption{\label{fig:vortexprofiles}Bond-pair-field magnitudes and local Josephson coupling near core 1 at $(15.5,7.5)$. Panels show (a) $A_i$, (b) $|\psi_i|$, (c) $|K_i|$ on the four axial cuts, and (d) $|K_i|$ on the two diagonal cuts. Panels (a,b) contain all six cuts; (c,d) use the same vertical range. Horizontal and vertical cuts follow the two nearest Fe rows and columns, at perpendicular offsets $\pm a/2$. Diagonal cuts run along $(1,1)$ and $(1,-1)$ through the core plaquette center, which contains no Fe site. The signed coordinate $s$ is measured along each cut from its closest point to the core; $|s|$ differs from radial distance on offset cuts. Markers are the original Fe-site values, spaced by $a$ on axial cuts and $\sqrt2a$ on diagonal cuts; lines connect successive values. Gray dashed lines give the clean value $A=|\psi|=0.05002$ in the stated energy units. Peaks in $|K|$ do not directly indicate a larger pair-field magnitude, and these magnitude profiles alone do not determine the coupling sign or phase winding.}
\end{minipage}
\end{widetext}

\renewcommand{\bibfont}{\small\setlength{\baselineskip}{9.8pt}}
\FloatBarrier
\bibliography{iron_QPI_JSTM_refs_v24_7}
\end{document}